\documentclass[11pt,a4paper,twoside]{article}
\usepackage{a4wide}
\usepackage[english]{babel}

\newcommand{\color}[1]{}

\usepackage{amsmath,amssymb,amsfonts,amscd}
\usepackage{dsfont}
\usepackage[mathscr]{eucal}

\newcommand {\anumber}  {}           
\newcommand {\eqip} [1] {#1}  

\newcommand {\eps} {\varepsilon}

\newcommand {\half} [1][1] {\ensuremath{\frac{#1}{2}}}

\newcommand {\ket} [1] {\ensuremath{\left| \!\!
                                      \begin{array}{c}
                                        #1 \\
                                      \end{array} \!\!
                                    \right>}}

\newcommand {\sups} [1] {\ensuremath{^{\textrm{\scriptsize #1}}}}

\newcommand {\mycite} [1] {\sups{\color{magenta}\cite{#1}}}

\DeclareMathOperator{\ad}{ad}

\renewcommand{\d}{\mbox{d}}

\renewcommand{\Re}{\mbox{Re}}

\newcommand {\ZZ}{\ensuremath{\mathds{Z}}}

\newcommand {\RR}{\ensuremath{\mathds{R}}}
\newcommand {\CC}{\ensuremath{\mathds{C}}}

\newcommand {\unit}{\ensuremath{\mathds{1}}}

\newcommand {\topk}  {\vec{k}}
\newcommand {\Bbk}   {b(\topk)}

\newcommand {\MS}    {\ensuremath{\mathcal{M}}}
\newcommand {\MTN}    {\MS_{TN}}
\newcommand {\ICS}   {\ensuremath{\mathcal{I}}}
\newcommand {\JCS}   {\ensuremath{\mathcal{J}}}
\newcommand {\KCS}   {\ensuremath{\mathcal{K}}}

\newcommand {\qBos}  {\mathsf{w}}
\newcommand {\qq}    {\mathsf{e}}
\newcommand {\qu}    {\mathsf{u}}
\newcommand {\qv}    {\mathsf{v}}
\newcommand {\qD}    {\mathsf{D}}
\newcommand {\qS}    {\mathsf{s}}

\newcommand {\QC}      {\mathcal{Q}}

  \newcommand {\ii}    {\imath}
\renewcommand {\ij}    {\jmath}
  \newcommand {\ik}    {\kappa}

\newcommand {\Jang} {\ensuremath{J}} 
\newcommand {\pauli} {\ensuremath{\sigma}}

\newcommand {\sfg}   {\mathsf{g}}
\newcommand {\nsusy} {N}
\newcommand {\nsusyms} {\mathcal{N}}

\newcommand {\sogen}{Y}

\newcommand {\Egamma} {\ensuremath{\overline{\gamma}}}

\newcommand {\ofr} {\ensuremath{e}}
\newcommand {\ofc} {\ensuremath{\theta}}

\begin{document}

\pagenumbering{arabic}

\title{  \bf Supercharges, Quantum States and Angular Momentum for  N=4 Supersymmetric  Monopoles                 }
\author{                                                                    \\
             Erik Jan de Vries\footnote{erikjandevries@hotmail.com}~ and Bernd J. Schroers\footnote{bernd@ma.hw.ac.uk}                                       \\
                                                                                        \\
         \it Department of Mathematics and Maxwell Institute for Mathematical Sciences, \\
         \it Heriot-Watt University, Edinburgh  EH14 4AS, UK                                \\
                                                                                        \\}
\date{      20 May 2010                                                         \\
             ~                                                                          \\
             \textsf{\small{{ EMPG-10-08}}}                                                }

\maketitle
\thispagestyle{empty}

\begin{abstract}
\noindent We revisit the moduli space approximation to the quantum mechanics of monopoles in $\nsusy=4$ supersymmetric Yang-Mills-Higgs theory with maximal symmetry breaking.
Starting with the observation that the set of fermionic zero-modes in $\nsusy=4$ supersymmetric Yang-Mills-Higgs theory can be viewed as two copies of the set of fermionic zero-modes in the $\nsusy=2$ version, we build a model to describe the quantum mechanics of $\nsusy=4$ supersymmetric monopoles, based on our previous paper\mycite{deVriesSchroers:susyQMMM} on the $\nsusy=2$ case,  in which this doubling of fermionic zero-modes is manifest throughout.
Our final picture extends the familiar result that quantum states  are described by differential forms on the moduli space and that the Hamiltonian operator is the Laplacian acting on forms. In particular, we derive a general expression for the  total angular momentum operator on the moduli space  which differs from the naive candidate by the adjoint action of the complex structures.  We also  express  all  the supercharges in terms of (twisted) Dolbeault operators and  illustrate our results by discussing, in some detail, the $\nsusy=4$ supersymmetric quantum dynamics of monopoles in a theory with gauge group $SU(3)$ broken to $U(1)\times U(1)$.
\end{abstract}




\section{Introduction}

Monopoles appear in certain classes of Yang-Mills-Higgs field theories with spontaneously broken symmetries\mycite{Hooft:Monopole,Polyakov:Monopole}. In this paper we use the moduli space approximation to study the quantum mechanics of monopoles in an $\nsusy=4$ supersymmetric version of Yang-Mills-Higgs theory. The work reported in this paper is an extension of our previous study of $\nsusy=2$ supersymmetric monopoles\mycite{deVriesSchroers:susyQMMM}, to which we refer for background material and further details.

The extension of the classical moduli space approximation\mycite{Manton:ScatteringBPSMon} to quantised monopole dynamics is essentially a collective coordinate quantisation, with the moduli spaces of monopoles providing sets of collective coordinates for single or several interacting monopoles. In the bosonic case\mycite{GibbonsManton:CQDynBPSMon,Schroers:QSBPSMon} quantum states are described by functions on   the moduli spaces, while in the supersymmetric case\mycite{Gauntlett:LowEnDymN=2Mon,Blum:SusyQMMonN=4} they are represented by differential forms on the moduli spaces. The Hamiltonian is the Laplace operator acting on these functions or forms. There exists an extensive literature, including reviews\mycite{WeinbergYi:MMonDynSusyDuality}, both on the justification of this approximation and on its application to particular cases.
However, what has been missing so far is an identification of physically important operators, such as the supercharges and the angular momentum operator, as (differential) operators naturally associated to the hyperk\"ahler geometry of the moduli spaces. In this paper we fill this gap by expressing the supercharges in terms of (twisted) Dolbeault operators, and the angular momentum operator in terms of a Lie derivative and the adjoint action of the complex structures. We derive our results from first principles, and illustrate them in particular examples.

Before we summarise the contents of this paper in more detail, we would like to comment on our expression for the angular momentum operator. The moduli spaces of monopoles inherit an $SO(3)$ action from the action of spatial rotations in the parent field theory.
The Lie derivatives associated to $SO(3)$ generators act on differential forms and thus provide geometrically natural candidates for the components of the angular momentum operator. However, geometrically natural objects defined on the moduli space - like differential forms or vector fields - are single-valued under a rotation by $2\pi$ and can therefore only carry integer spins with respect to an angular momentum operator defined in this way. This means that a quantum mechanical model in which states are represented by forms on the moduli space can only describe the half-integer spin states of the original supersymmetric field theory if the angular momentum operator is defined differently. The observation that the correct expression for the angular momentum operator involves not only the Lie derivative described above but also extra terms was first made in a particular case by Bak, Lee and Yi\mycite{BakLeeYi:quantumdyons}.
In our previous paper\mycite{deVriesSchroers:susyQMMM} on the $\nsusy = 2$ supersymmetric theory we proposed a general expression for the angular momentum operator, where the extra terms are expressed in terms of the adjoint action of the complex structures on the moduli spaces. In this paper we derive this expression for the angular momentum operator from considerations in the field theory. Then we show that it has the correct commutation relations with all of the supercharges and that it also gives rise to the expected spin contents of supermultiplets in the context of the $\nsusy=4$ supersymmetric theory.  Our general expression for the angular momentum operator agrees with the expression proposed by Bak, Lee and Yi\mycite{BakLeeYi:quantumdyons} in the particular case of charge-$(1,1)$ monopoles in $SU(3)$ gauge theory spontaneously broken to $U(1)\times U(1)$.

We begin, in section~\ref{sectN4BPSMon}, with a short review of BPS monopoles, the zero-modes of the theory and the effective Lagrangian in the moduli space approximation\mycite{Gauntlett:LowEnDymN=2Mon,Blum:SusyQMMonN=4}. We introduce a quaternionic notation and  use it to explain how the complex structures act on both the bosonic and fermionic zero-modes.  We show that both bosonic and fermionic zero-modes can be written as quaternion-valued functions on $\RR^3$, and use this common notation to compare the angular momentum operator for bosonic and fermionic zero-modes. The difference between the two angular momentum operators can be expressed in terms of the complex structures, leading us to the promised expression for the components of the  angular momentum operator for fermionic zero-modes as the geometrically natural rotation generators acting on forms plus a correction term involving the complex structures.

In section~\ref{sectN4FZMFMS}, we discuss the effective action of the $\nsusy = 4$ supersymmetric model and its quantisation. A key reference here is the review by Weinberg and Yi\mycite{WeinbergYi:MMonDynSusyDuality}. As in that review and other papers on this subject, we arrive at the conclusion that quantum states of supersymmetric monopoles are differential forms on the moduli space. However, we choose a slightly different quantisation prescription which allows us to identify two independent (but equivalent) sets of fermionic zero-modes, both of which are isomorphic to the set of fermionic zero-modes in the $\nsusy=2$ supersymmetric theory. This duplication of the zero-modes is a direct result of the doubling of the supersymmetry. In the quantised theory,  we identify the two sets of fermionic zero-modes  with, respectively,  the set of  anti-holomorphic forms and the set of holomorphic forms, thus obtaining a natural extension of the quantisation procedure for $\nsusy=2$ supersymmetric monopoles, where quantum states of monopoles are described by anti-holomorphic forms on the moduli space.
We construct the differential operators corresponding to all the supercharges, and identify them as (twisted) Dolbeault operators and their adjoints.

We then go on, in section~\ref{sectAngMom}, to show that our considerations in section~\ref{sectN4BPSMon} lead to  the expression  for the total angular momentum operator as  the  differential operator on the moduli space already proposed in our paper on $\nsusy = 2$ supersymmetric monopoles\mycite{deVriesSchroers:susyQMMM}.  We check that it has the required commutation relations with the operators representing the supersymmetry charges.

In sections~\ref{sectEx1} and \ref{sectEx2} the general results of this paper are illustrated for the cases of single (charge-$1$) monopoles and charge-$(1,1)$ monopoles in $\nsusy = 4$ supersymmetric Yang-Mills-Higgs theory with gauge group $SU(3)$ spontaneously broken to $U(1)\times U(1)$. Expressing the supercharges explicitly in terms of coordinates on the moduli spaces we construct the supermultiplets and, using equally explicit expressions for our angular momentum operator, we check that they have the expected spin contents.
We show that the hyperk\"ahler forms on the moduli space are singlets under the action of the total angular momentum operator, even though they form a triplet under the geometrically natural $SO(3)$ action on the moduli space. We also check that the unique harmonic and square-integrable two-form on the relative moduli space of monopoles of charge-$(1,1)$, which plays an important role in checking S-duality for $\nsusy = 4$ supersymmetric Yang-Mills-Higgs theory, is a singlet under the proposed angular momentum operator,  as required by uniqueness.
Finally, we give a brief outlook onto possible applications of our results in section~\ref{outlook}.

\section{$\nsusy=4$ supersymmetric BPS monopoles}
\label{sectN4BPSMon}

\subsection{$\nsusy=4$ supersymmetric Yang-Mills-Higgs theory}
We study Yang-Mills-Higgs models with gauge group $SU(n)$ on four-dimensional Minkowski space with metric $\eta$ of signature $(+,-,-,-)$ . The gauge field is denoted $A_\mu$, the associated covariant derivative is $D_\mu = \partial_\mu - e \ad A_\mu$, where $-e$ is the coupling constant, and the corresponding curvature is $F_{\mu\nu} = \partial_\mu A_\nu - \partial_\mu A_\nu - e [A_\mu,A_\nu]$.
Using $\cdot$ to denote an inner product on the Lie algebra $su(n)$, and $|| \ \  ||$ for the associated norm, the $\nsusy=4$ supersymmetric Lagrangian that we are interested in is the following, adapted from Blum\mycite{Blum:SusyQMMonN=4} and Osborn\mycite{Osborn:TopChSpin}:
\begin{align}
L 
 & \ = \ \int \d^3 x \Big( - \frac{1}{4} F^{\mu\nu} \cdot F_{\mu\nu} + \half D_{\mu} S_{\ii} \cdot D^{\mu} S_{\ii} + \half D_{\mu} P_{\ij} \cdot D^{\mu} P_{\ij} \nonumber \\
 & \qquad\qquad\quad - \frac{e^2}{4} \left( || \left[ S_{\ii}, S_{\ij} \right] ||^2 + 2 || \left[ S_{\ii}, P_{\ij} \right] ||^2 + || \left[ P_{\ii}, P_{\ij} \right] ||^2 \right) \nonumber \\
 & \qquad\qquad\quad + \half[i] \overline{\psi}_r \cdot \gamma^{\mu} D_{\mu} \psi_r + \half[e] \overline{\psi}_r \cdot \left( \alpha^{\ii}_{rs} \, \ad S_{\ii} - i \beta^{\ij}_{rs} \gamma^5 \, \ad P_{\ij} \right) \psi_s\Big)\eqip{.} \label{LN4}
\end{align}
Here $S_{\ii}$ are three scalar fields, $P_{\ij}$ are three pseudo-scalar fields, and $\psi_r$ are four Majorana spinors, all in the adjoint representation of the gauge group.
The indices have the following ranges: $\ii, \ij, \ldots \in \{ 1,2,3 \}$ and $r, s, \ldots \in \{1,2,3,4\}$. We take the $\gamma$-matrices to satisfy $\{ \gamma_{\mu}, \gamma_{\nu} \} = 2 \eta_{\mu\nu}$ and the chiral operator $\gamma_5$ is defined by $\gamma_5 = i \gamma_0 \gamma_1 \gamma_2 \gamma_3$. The $4\times4$ matrices
$\alpha^{\ii}$ and $\beta^{\ij}$ are real and anti-symmetric, and satisfy
\begin{align}
  \left[ \alpha^{\ii}, \alpha^{\ij} \right] \ = \ & -2 \eps^{\ii\ij\ik} \alpha^{\ik}\eqip{,} &
  \left\{ \alpha^{\ii}, \alpha^{\ij} \right\} \ = \ & -2 \delta^{\ii\ij} \unit_4\eqip{,} \nonumber \\
  \left[ \beta^{\ii}, \beta^{\ij} \right] \ = \ & -2 \eps^{\ii\ij\ik} \beta^{\ik}\eqip{,} &
  \left\{ \beta^{\ii}, \beta^{\ij} \right\} \ = \ & -2 \delta^{\ii\ij} \unit_4\eqip{,} \nonumber \\
  \left[ \alpha^{\ii}, \beta^{\ij} \right] \ = \ & \phantom{-} 0\eqip{.} \anumber
\end{align}
They generate commuting $so(3)$ subalgebras of $so(4) = so(3) \oplus so(3)$.

The Majorana condition is the requirement that the spinors $\psi_r$ are invariant under the (anti-linear) charge conjugation operation
\begin{align}
{\cal C} (\psi_r) \ = \ C \gamma_0 \psi_r^*\eqip{,} \label{chargecon}
\end{align}
where $C$ is the charge conjugation matrix (we will choose a convenient representation below).

The Lagrangian \eqref{LN4} can be derived from an $\nsusy=1$ supersymmetric model in ten dimensions, via dimensional reduction\mycite{BrinkSchwSch:SusyYMT,DAddaEtAl:SuSyMMD}. The rotational symmetry of the six extra dimensions reduces to an $SU(4)$ internal symmetry of the Lagrangian in four dimensions. This $SU(4)$ internal group, which is the double cover of $SO(6)$, is generated by the six $so(4)$ generators $\alpha^{\ii}$ and $\beta^{\ij}$ and the nine anti-commutators $\{\alpha^{\ii}, \beta^{\ij}\}$.

For many of the calculations it is convenient to have explicit representations of the matrices $\alpha^{\ii}$, $\beta^{\ij}$, the $\gamma$-matrices and the charge conjugation matrix $C$. We will use the following:
\begin{align}
\alpha^1 & \ = \ \left(
                           \begin{array}{cccc}
                             0 & 0 & 0 & 1 \\
                             0 & 0 & 1 & 0 \\
                             0 & -1 & 0 & 0 \\
                             -1 & 0 & 0 & 0 \\
                           \end{array}
                         \right) &
\beta^1 
                 & \ = \ \left(
                           \begin{array}{cccc}
                             0 & 0 & 0 & -1 \\
                             0 & 0 & 1 & 0 \\
                             0 & -1 & 0 & 0 \\
                             1 & 0 & 0 & 0 \\
                           \end{array}
                         \right) \nonumber \displaybreak[2] \\
\alpha^2 & \ = \ \left(
                           \begin{array}{cccc}
                             0 & 0 & -1 & 0 \\
                             0 & 0 & 0 & 1 \\
                             1 & 0 & 0 & 0 \\
                             0 & -1 & 0 & 0 \\
                           \end{array}
                         \right) &
\beta^2 
                 & \ = \ \left(
                           \begin{array}{cccc}
                             0 & 0 & -1 & 0 \\
                             0 & 0 & 0 & -1 \\
                             1 & 0 & 0 & 0 \\
                             0 & 1 & 0 & 0 \\
                           \end{array}
                         \right) \nonumber \displaybreak[2] \\
\alpha^3 & \ = \ \left(
                           \begin{array}{cccc}
                             0 & 1 & 0 & 0 \\
                             -1 & 0 & 0 & 0 \\
                             0 & 0 & 0 & 1 \\
                             0 & 0 & -1 & 0 \\
                           \end{array}
                         \right) &
\beta^3 
                 & \ = \ \left(
                           \begin{array}{cccc}
                             0 & 1 & 0 & 0 \\
                             -1 & 0 & 0 & 0 \\
                             0 & 0 & 0 & -1 \\
                             0 & 0 & 1 & 0 \\
                           \end{array}
                         \right)\eqip{.} \label{repalphabeta}
\end{align}
A convenient representation for the $\gamma$-matrices is given by
\begin{align}
  \gamma_0 & \ = \ \left(
                                \begin{array}{cc}
                                  0 & \unit_2 \\
                                  \unit_2 & 0 \\
                                \end{array}
                              \right)\eqip{,} &
  \gamma_i & \ = \ \left(
                                \begin{array}{cc}
                                  -i \sigma_i & 0 \\
                                  0 & i \sigma_i \\
                                \end{array}
                              \right)\eqip{,} &
  \gamma_5 & \ = \ i \left(
                                \begin{array}{cc}
                                  0 & \unit_2 \\
                                  -\unit_2 & 0 \\
                                \end{array}
                              \right)\eqip{.} \label{gammaMatr}
\end{align}
Finally, for the charge conjugation matrix we choose
\begin{align}
  C & \ = \ \left(
                                \begin{array}{cc}
                                  i \pauli_2 & 0 \\
                                  0 & -i \pauli_2 \\
                                \end{array}
                              \right)\eqip{.} \label{chargeConjMatr}
\end{align}

\subsection{BPS monopoles}
Monopoles may appear in Yang-Mills-Higgs theory when the gauge symmetry is spontaneously broken. This is accomplished by imposing appropriate boundary conditions on the (pseudo-) scalar fields at infinity. Before we come to these boundary conditions, we first have a look at the kinetic and potential energy for the $\nsusy = 4$ supersymmetric model, which are given by
\begin{align}
K & \ = \ \int \d^3 x \ \Big( || \half F_{0i} ||^2 + \half || D_{0} S_{\ii} ||^2 + \half || D_{0} P_{\ij} ||^2 + \half[i] \overline{\psi}_r \cdot \gamma^{0} D_{0} \psi_r \Big)\eqip{,} \anumber \\
V & \ = \ \int \d^3 x \ \Big( \ \frac{1}{4} || F_{ij}||^2 + \half || D_{i} S_{\ij} ||^2 + \half || D_{i} P_{\ij} ||^2 \nonumber \\
  & \qquad\qquad\qquad + \frac{e^2}{4} \left( || \left[ S_{\ii}, S_{\ij} \right] ||^2 + 2 || \left[ S_{\ii}, P_{\ij} \right] ||^2 + || \left[ P_{\ii}, P_{\ij} \right] ||^2 \right) \nonumber \\
  & \qquad\qquad\qquad + \half[i] \overline{\psi}_r \cdot \gamma_{i} D_{i} \psi_r - \half[e] \overline{\psi}_r \cdot \left( \alpha^{\ii}_{rs} \, \ad S_{\ii} - i \beta^{\ij}_{rs} \gamma^5 \, \ad P_{\ij} \right) \psi_s \Big)\eqip{.} \label{VN4}
\end{align}
The potential energy involves the norms of commutators of the scalar fields $S_{\ii}$ and $P_{\ij}$. For finite energy configurations these commutators therefore necessarily vanish at spatial infinity. This is possible even if  $S_{\ii}$ or $P_{\ij}$  are non-vanishing at spatial infinity,  in which case the gauge symmetry  is spontaneously broken. A parity-invariant vacuum  requires that the vacuum expectation values of the pseudoscalar fields $P_{\ij}$ are zero.

In order to break the gauge symmetry without breaking parity,  we impose the boundary conditions $\lim_{r\to\infty} \left( P_{\ii} \cdot P_{\ii} \right) = 0$ and  $\lim_{r\to\infty} \left( S_{\ii} \cdot S_{\ii} \right) = a^2, a \neq 0$. In this paper we only consider the case where the values of $(S_1,S_2,S_3)$ at spatial infinity break the gauge symmetry to a maximal torus; specifically, $SU(n)$ is broken to $U(1)^{n-1}$.
We can use the $SU(4)$ internal symmetry of the Lagrangian to pick a direction of the vacuum expectation value of $(S_1,S_2,S_3)$. In the following, we will assume that only the scalar field $S_3$ has a non-zero vacuum expectation value.  We set the remaining Higgs fields to zero everywhere, so that the bosonic sector is effectively the same as a non-supersymmetric Yang-Mills-Higgs theory with a single adjoint Higgs field $S_3$.

With the above assumptions on the behaviour of the scalar and pseudo-scalar fields at spatial infinity,
the bosonic finite energy configurations are classified by a topological charge $\vec{k}\in \ZZ^{n-1}$, which measures the magnetic charge of the configuration. The magnetic charge is the $su(n)$-valued coefficient of the
leading $1/r^2$ term of the non-Abelian magnetic field near spatial infinity. In practice, one picks a direction in space and reads off the magnetic charge  and the asymptotic value of the Higgs field from a large-$r$ expansion of the magnetic field and the Higgs field $S_3$ in that direction respectively.
One may choose both the  magnetic charge and the asymptotic value of $S_3$ to lie in a fixed Cartan subalgebra of $su(n)$.  The  components of $\vec{k}$ are the coefficients in the expansion of the magnetic charge in a basis of co-roots which have a positive inner product with the asymptotic value of $S_3$. Further details about the magnetic charges and their topological interpretation can be found in the classic paper by Goddard, Nuyts  and Olive\mycite{GNO:GTMC} or, in more mathematical language, in the work of  Taubes\mycite{Taubes:magcharges}.

For a given charge, the potential energy is bounded from below by the Bogomol'nyi bound\mycite{Bogomolnyi:Stability}, $V \geq \frac{4\pi a}{e}\Bbk$, where $\Bbk$ is a positive, real function of that topological charge.
BPS monopoles are defined to be the configurations of minimal energy in sectors with non-vanishing topological charge. They satisfy one of the Bogomol'nyi equations
\begin{align}
  B_i \equiv \half \epsilon_{ijk} F^{jk} = \pm D_i S_3\eqip{,} \label{Bogeq}
\end{align}
with the sign depending on the sign of the topological charge (i.e. the sign of the components $k_i$ of the topological charge $\vec{k}$, which are either all non-negative or all non-positive; see also section~\ref{sectQuatNot}).

\subsection{The Dirac operator in $\nsusy=4$ supersymmetric Yang-Mills-Higgs theory}
\label{sectR4}
To minimise the fermionic part of the potential energy \eqref{VN4}, the quartet of Majorana spinors $(\psi_1,\psi_2,\psi_3,\psi_4)$ must be a zero-mode of the Dirac operator coupled to the monopole background, i.e. it must satisfy
\begin{align}
  \gamma_0 \gamma_i D_{i} \psi_r + i e \gamma_0 \alpha^3_{rs} \ad S_3 \psi_s & \ = \ 0\eqip{.} \label{diracFZMN4}
\end{align}
In studying this Dirac operator it is convenient to think  of it as a  Dirac operator in a four-dimensional Euclidean space, with all fields independent of the newly introduced fourth Euclidean dimension: $\partial_4 \equiv 0$.

To make this explicit, we  define
\begin{align}
W_i & \ = \ A_i\eqip{,} &
W_4 & \ = \ S_3\eqip{.} \label{defW}
\end{align}
Then we  think of $W_{\underline{i}}$,  with the underlined indices $\underline{i}$ running  from 1 to 4, as a connection on Euclidean $\RR^4$. The corresponding covariant derivatives are defined by
\begin{align}
  D_{\underline{i}} & \ = \ \partial_{\underline{i}} \, - e \ad W_{\underline{i}}\eqip{.} \label{defD}
\end{align}
We  also define Euclidean $\Egamma$-matrices via
\begin{align}
\Egamma_i & \ = \ \gamma_0 \gamma_i\eqip{,} &
\Egamma_4 & \ = \ \gamma_0\eqip{.} \label{euclidGammaMatr}
\end{align}
They satisfy the Euclidean Clifford relations $\{ \Egamma_{\underline{i}} , \Egamma_{\underline{j}} \} = 2 \delta_{\underline{ij}}$. The  Euclidean version of the chirality operator is given by  $\Egamma_5 = \Egamma_1 \Egamma_2 \Egamma_3 \Egamma_4$.
Using representation \eqref{gammaMatr} for the space-time $\gamma$-matrices, the Euclidean $\Egamma$-matrices become
\begin{align}
  \Egamma_i & \ = \ \left(
                                \begin{array}{cc}
                                  0 & i \sigma_i \\
                                  -i \sigma_i & 0 \\
                                \end{array}
                              \right)\eqip{,} &
  \Egamma_4 & \ = \ \left(
                                \begin{array}{cc}
                                  0 & \unit_2 \\
                                  \unit_2 & 0 \\
                                \end{array}
                              \right)\eqip{,} &
  \Egamma_5 & \ = \ \left(
                                \begin{array}{cc}
                                  -\unit_2 & 0 \\
                                  0 & \unit_2\\
                                \end{array}
                              \right)\eqip{.} \label{euclidGammaMatrRepr}
\end{align}
In terms of the Euclidean $\Egamma$-matrices  and in the temporal gauge $A_0 = 0$, the Dirac equation \eqref{diracFZMN4} becomes
\begin{align}
\Egamma_i D_{i} \psi_r - i \Egamma_4 \alpha^3_{rs} D_4 \psi_s & \ = \ 0\eqip{.} \label{diracFZMN4a}
\end{align}

It turns out that the equation \eqref{diracFZMN4a} for a quartet of Majorana fermions $\psi_r$ can be separated into two uncoupled equations for a pair of Dirac fermions. This observation is fundamental for the remainder of this paper.
 We  explain it  now  using the explicit representation \eqref{repalphabeta} of $\alpha^3$. Defining  two Dirac fermions as the linear combinations
\begin{align}
\xi^+ & \ = \ \psi_1-i\psi_2\eqip{,} &
\xi^- & \ = \ \psi_3-i\psi_4\eqip{,} \label{Diracspinors}
\end{align}
it is straightforward to check that, if the quartet of Majorana spinors $\psi_r$ satisfy the Dirac equation \eqref{diracFZMN4a}, then $\xi^+$ and $\xi^-$ satisfy the two uncoupled Dirac  equations
\begin{align}
\Egamma_{\underline{i}} D_{\underline{i}} \, \xi^+ & \ = \ 0\eqip{,} &
\Egamma_{\underline{i}} D_{\underline{i}} \, \xi^- & \ = \ 0\eqip{.} \label{diracFZMN4uncoupled}
\end{align}
Thus, in order  to minimise the fermionic part of the potential energy,  the  Dirac fermions $\xi^+$ and $\xi^-$ both have to  obey the same Dirac equation as the fermion field in the $\nsusy = 2$ supersymmetric model\mycite{Gauntlett:LowEnDymN=2Mon,deVriesSchroers:susyQMMM}.

Conversely, suppose we have solutions $\xi^+$ and $\xi^-$ of the Dirac equations \eqref{diracFZMN4uncoupled}. Then we can reconstruct a quartet of Majorana fermions $\psi_r$ which solve the Dirac equation \eqref{diracFZMN4a} as follows. First, we use the charge conjugation operator \eqref{chargecon} to define the two projection operators $P_+$ and $P_-$ via
\begin{align}
  P_{\pm} & \ = \ \half ( 1 \pm \mathcal{C} )\eqip{.} \anumber
\end{align}
They add up to the identity operator, $P_+ + P_- = 1$. Since $\mathcal{C}^2 = 1$, we have $\mathcal{C} \circ P_{\pm} = \pm P_{\pm}$, which means that the operator $P_+$ projects onto Majorana spinors and the operator $P_-$ projects onto spinors which are mapped to their negative under charge conjugation. Since $\mathcal{C}$ is anti-linear, the four non-vanishing spinors
\begin{align}
  \psi_1 & \ = \ P_+ \xi^+\eqip{,} &
  \psi_2 & \ = \ i P_- \xi^+\eqip{,} &
  \psi_3 & \ = \ P_+ \xi^-\eqip{,} &
  \psi_4 & \ = \ i P_- \xi^- \label{MajSpinors}
\end{align}
are Majorana spinors. They satisfy the Dirac equation \eqref{diracFZMN4} if $\xi^+$ and $\xi^-$ satisfy the uncoupled equations \eqref{diracFZMN4a}. It is not difficult to check that the relations \eqref{MajSpinors} and \eqref{Diracspinors} are inverses of each other, so that we have established
a bijection between a pair of Dirac spinors solving equations \eqref{diracFZMN4uncoupled} and a quartet of Majorana spinors solving the Dirac equation \eqref{diracFZMN4a}.

\subsection{Bosonic zero-modes}
\label{sectQuatNot}
The bosonic zero-modes of the $\nsusy = 4$ supersymmetric monopole are exactly the same as those of the purely bosonic monopole. The standard approach for dealing with the bosonic zero-modes is to consider moduli spaces, which are spaces of gauge equivalence classes of solutions of the Bogomol'nyi equations \eqref{Bogeq} for a given topological charge $\vec{k}$. The points in a given moduli space thus parameterise all minimal energy configurations for the given topological charge $\vec{k}\in \ZZ^{n-1}$.
Moduli spaces are non-empty iff the components $k_1,\ldots, k_{n-1}$ are either all positive or all negative, as was first shown by Weinberg\mycite{Weinberg:ParamCount}  who also  computed the dimension of the moduli space to be $4k$, where $k=|k_1 + \ldots + k_{n-1}|$. It was subsequently shown that the Riemannian metric which the moduli spaces inherit from the kinetic energy of the field theory is, in fact, hyperk\"ahler. For details we refer reader to the book by Atiyah and Hitchin\mycite{AtiyahHitchin}.

Tangent vectors to the moduli space represent gauge equivalence classes of  bosonic zero-modes. We use the notation  $\delta W_{\underline i}$  for a generic bosonic zero-mode.
As explained in our previous paper\mycite{deVriesSchroers:susyQMMM}, a quaternionic language is convenient for studying such zero-modes. In the notation of that paper, with  $\qq_1$, $\qq_2$ and $\qq_3$  denoting the unit imaginary quaternions and $\qq_4 = 1$ the identity quaternion, the quaternion algebra is determined by
\begin{align}
 \label{quatalg}
\qq_i\qq_j & \ = \ -\delta_{ij}\qq_4 + \epsilon_{ijk}\qq_k\eqip{,}
\end{align}
with summation over repeated indices assumed. Quaternionic conjugation is given by $\overline{\qq}_i = -\qq_i$ while $\overline{\qq}_4 = \qq_4$.
We now define
\begin{align}
\qBos & \ = \ \overline{\qq}_{\underline{i}} \, \delta W_{\underline{i}} \label{bzmq},
\end{align}
so that the linearised Bogomol'nyi equations and Gauss' law  can be expressed in the single quaternionic equation
\begin{align}
\label{bosonzero}
 \qD \qBos & \ = \ 0\eqip{,}
\end{align}
where $\qD$ is the Dirac operator
\begin{align}
  \qD & \ = \ \qq_{\underline{j}} D_{\underline{j}}\eqip{.} \label{smallDirac}
\end{align}

\subsection{Fermionic zero-modes}
Since the Dirac equations \eqref{diracFZMN4uncoupled} are already linear, the fermionic zero-modes of $\nsusy = 4$ supersymmetric monopoles are their solutions $\xi^+$ and $\xi^-$. The Dirac equations \eqref{diracFZMN4uncoupled} have the same form as the Dirac equation in the $\nsusy=2$ supersymmetric theory, and therefore we conclude that the fermionic zero-modes of the $\nsusy = 4$ supersymmetric model correspond to two independent sets of fermionic zero-modes, both isomorphic to the set of fermionic zero-modes in the $\nsusy = 2$ supersymmetric model\mycite{Gauntlett:LowEnDymN=2Mon,deVriesSchroers:susyQMMM}. We now briefly review the form of the Dirac zero-modes and explain their relation to bosonic zero-modes.

With our conventions, the Dirac operator entering the Dirac equations \eqref{diracFZMN4uncoupled} can be written
\begin{align}
  \Egamma_{\underline{i}} D_{\underline{i}} & \ = \
    \begin{pmatrix}  0 &  \qD^\dagger \\
      \qD & 0 \end{pmatrix}\eqip{.} \label{bigDirac}
\end{align}
where $\qD$ is the Dirac operator \eqref{smallDirac} and we have identified the imaginary unit quaternions with the Pauli matrices via
\begin{align}
\label{quatrep}
  \qq_i & \ = \ -i\sigma_i\eqip{,} &
  \qq_4 & \ = \ \unit_2\eqip{.} \anumber
\end{align}
It follows from an index theorem and a vanishing theorem  for $\qD$, both  due to Taubes\mycite{Taubes:index},
 that
the solutions to equations \eqref{diracFZMN4uncoupled}
are of the form
\begin{align}
\label{4two2}
\xi^\pm & \ = \begin{pmatrix}
     \eta^\pm \\
     0 \\
  \end{pmatrix} \eqip{,} \anumber
\end{align}
which satisfy $\Egamma_5 \xi^{\pm} = - \xi^{\pm}$, where $\eta^+$ and $\eta^-$ are two-component spinors satisfying $\qD\eta^\pm=0$.
From \eqref{MajSpinors} we find that the corresponding quartet of  Majorana spinors consists of
\begin{align}
\label{Majoranaconst}
  \psi_{1} & \ = \ \phantom{i}P_+ \left(
       \begin{array}{c}
         \eta^+ \\
         0 \\
       \end{array} \right) \ = \ \half \left(
                             \begin{array}{c}
                               \eta^+ \\
                               -i \pauli_2 (\eta^+)^* \\
                             \end{array} \right)\eqip{,} \nonumber \\
\psi_{2} & \ = \ i P_- \left(
       \begin{array}{c}
         \eta^+ \\
         0 \\
       \end{array} \right) \ = \ \half[i] \left(
                             \begin{array}{c}
                               \eta^+ \\
                               i \pauli_2 (\eta^+)^* \\
                             \end{array} \right)\eqip{,} \nonumber \\
\psi_{3} & \ = \ \phantom{i}P_+ \left(
       \begin{array}{c}
         \eta^- \\
         0 \\
       \end{array}\right) \ = \ \half \left(
                             \begin{array}{c}
                               \eta^- \\
                               -i \pauli_2 (\eta^-)^* \\
                             \end{array}
                           \right)\eqip{,} \nonumber \\
\psi_{4} & \ = \ i P_- \left(
       \begin{array}{c}
         \eta^- \\
         0 \\
       \end{array} \right) \ = \ \half[i] \left(
                             \begin{array}{c}
                               \eta^- \\
                               i \pauli_2 (\eta^-)^* \\
                             \end{array}
                           \right)\eqip{.}
\end{align}

We may use the fact that bosonic zero-modes $\qBos$ satisfy the Dirac equation \eqref{bosonzero} to relate the fermionic zero-modes to the bosonic ones. Using \eqref{quatrep} to write a bosonic zero-mode as a $2 \times 2$ matrix, it is clear that  $\qD \qBos \chi^\pm =0$ for any constant two-component spinors $\chi^\pm$. Therefore
\begin{align}
\xi^\pm   \ = \
             \begin{pmatrix}
               \qBos \, \chi^\pm \\
               0 \\
             \end{pmatrix}  \label{fermbos}
\end{align}
are fermionic zero-modes, i.e. $\xi^+$ and $\xi^-$ satisfy the Dirac equations \eqref{diracFZMN4uncoupled}, for any $\chi^+$ or $\chi^-$. As we shall see below, the various possibilities for $\chi^+$ and $\chi^-$ are related through the action of the complex structures\mycite{Gauntlett:LowEnDymN=2Mon}, so that we only need to pick a single fixed $\chi^+$ and a single fixed $\chi^-$ to parameterise all fermionic zero-modes. A convenient choice is given by
\begin{align}
  \chi^+ & \ = \ \left(
    \begin{array}{c}
      1 \\
      0 \\
    \end{array}
  \right)\eqip{,} &
  \chi^- & \ = \ \left(
    \begin{array}{c}
      0 \\
      1 \\
    \end{array}
  \right)\eqip{.} \label{choicechiN4}
\end{align}
With this choice  we have $-i\sigma_2(\chi^+)^* =\chi^-$ and $-i\sigma_2(\chi^-)^* =-\chi^+$. Using these relations, and  the fact that $e_2=-i\sigma_2$ is real while $e_1=-i\sigma_1$ and $e_3=-i\sigma_3$ are imaginary, the quartet of Majorana fermions \eqref{Majoranaconst}  can be written as
\begin{align}
  \psi_{1} & \ = \ \half \left( \begin{array}{r} \phantom{-} \qBos  \chi^+ \\ \phantom{-} \qBos  \chi^- \end{array} \right) \eqip{,} &
  \psi_{2} & \ = \ \half[i] \left( \begin{array}{r} \phantom{-} \qBos  \chi^+ \\ -\qBos  \chi^- \end{array} \right) \eqip{,} \nonumber \\
  \psi_{3} & \ = \ \half \left( \begin{array}{r} \phantom{-} \qBos  \chi^- \\ -\qBos  \chi^+ \end{array} \right) \eqip{,} &
  \psi_{4} & \ = \ \half[i] \left( \begin{array}{r} \phantom{-} \qBos  \chi^- \\ \phantom{-} \qBos  \chi^+ \end{array} \right) \eqip{.} \label{bosomaj}
\end{align}
Together,  these expressions provide a map from a bosonic zero-mode $\qBos$ to a quartet of Majorana zero-modes $(\psi_1, \psi_2, \psi_3, \psi_4)$.


\subsection{The action of the complex structures}
\label{sectComplStructAct}
As discussed above,  both the combination of the Bogomol'nyi equations and Gauss' law \eqref{bosonzero} and the Dirac equations \eqref{diracFZMN4a} or \eqref{diracFZMN4uncoupled} can be viewed as equations on Euclidean 4-space, $\RR^4$, with all fields independent of the fourth coordinate and the covariant derivatives defined by \eqref{defD}.
Euclidean $\RR^4$  has a natural hyperk\"ahler structure,  and it turns out that the
associated complex structures all act naturally on both the bosonic and the fermionic zero-modes discussed in the previous sections. We  now exhibit these actions.

Using the quaternionic notation introduced in section~\ref{sectQuatNot}, the three complex structures that make up the hyperk\"ahler structure act on the bosonic zero-modes simply by quaternionic multiplication:
\begin{align}
  \ICS_i(\qBos) & \ = \ - \qBos \, \qq_i\eqip{.} \label{hyperaction}
\end{align}
This action is a linear representation of the complex structures on the set of bosonic zero-modes, i.e. it manifestly maps a solution of the bosonic zero-mode equation \eqref{bosonzero} to another solution. As reviewed in our previous paper\mycite{deVriesSchroers:susyQMMM}, it also preserves the kinetic energy and hence the metric on the moduli space.

The quartet of Majorana spinors  associated to a bosonic zero-mode $\qBos$ via \eqref{bosomaj} is permuted under the action of the complex structures  \eqref{hyperaction} on $\qBos$ as follows
\begin{align}
 \ICS_1: \ (\psi_{1},\psi_{2},\psi_{3},\psi_{4}) & \ \mapsto \ (\phantom{-}\psi_{4},-\psi_{3},\phantom{-}\psi_{2},-\psi_{1})\eqip{,} \nonumber \\
 \ICS_2: \ (\psi_{1},\psi_{2},\psi_{3},\psi_{4}) & \ \mapsto \ (-\psi_{3},-\psi_{4},\phantom{-}\psi_{1},\phantom{-}\psi_{2})\eqip{,} \nonumber \\
 \ICS_3: \ (\psi_{1},\psi_{2},\psi_{3},\psi_{4}) & \ \mapsto \ (\phantom{-}\psi_{2},-\psi_{1},-\psi_{4},\phantom{-}\psi_{3})\eqip{,} \label{hypermaj}
\end{align}
 and the  pair of Dirac spinors corresponding to this quartet of Majorana spinors  via \eqref{Diracspinors} is  acted on according to
\begin{align}
 \ICS_1: \ (\xi^+,\xi^-) & \ \mapsto \ (\phantom{-}i\xi^-,\phantom{-}i\xi^+)\eqip{,} \nonumber \\
 \ICS_2: \ (\xi^+,\xi^-) & \ \mapsto \ (-\phantom{i}\xi^-, \phantom{-i}\xi^+)\eqip{,} \nonumber \\
 \ICS_3: \ (\xi^+,\xi^-) & \ \mapsto \ (\phantom{-}i\xi^+, -i\xi^-)\eqip{.} \label{hyperdirac}
\end{align}
The maps \eqref{hypermaj} form a linear representation of the action of the complex structures on the set of Majorana zero-modes (i.e. they map a solution of the Dirac equation \eqref{diracFZMN4a} into another solution). Similarly, the maps \eqref{hyperdirac} form a linear representation of the action of the complex structures on the set of Dirac zero-modes (mapping solutions of the Dirac equation \eqref{diracFZMN4uncoupled} into each other).
We thus have actions of the complex structures on both bosonic and fermionic zero-modes. The bijection \eqref{bosomaj} between bosonic and fermionic zero-modes, using our choice \eqref{choicechiN4} for $\chi^\pm$, is an intertwiner of these actions, i.e. it commutes with them. 

Note that the action of the complex structures on the Dirac zero-modes \eqref{hyperdirac} crucially depends on the presence of {\em two} Dirac  spinors $\xi^\pm$. In the $\nsusy=2$ supersymmetric model there is only one Dirac fermion and, as a result, we were not able to define {\em linear} actions of all of the complex structures  on the fermionic zero-modes\mycite{deVriesSchroers:susyQMMM}. Instead we were forced to define an action of the complex structures which is linear for one of them ($\ICS_3$), but anti-linear for the other two ($\ICS_1$ and $\ICS_2$).

As we will see below, with our  choice \eqref{choicechiN4} of  $\chi^\pm$, we are able to identify the Dirac zero-modes $\xi^+$ with anti-holomorphic forms with respect to $\ICS_3$ on the moduli space, and the Dirac zero-modes $\xi^-$ with holomorphic forms. To reflect the special role played by $\ICS_3$ in our notation, we sometimes denote the complex structures as follows
\begin{align}
\ICS & \ = \ \ICS_3\eqip{,} &
\JCS & \ = \ \ICS_1\eqip{,} &
\KCS & \ = \ \ICS_2\eqip{.} \label{rename}
\end{align}

It is worth commenting on the role played by our choice of the constant two-component spinors $\chi^\pm$ in equations \eqref{choicechiN4}. If, instead of the standard basis of $\CC^2$, we had picked another orthonormal basis $\tilde{\chi}^\pm$, this basis would necessarily be related to our chosen basis $\chi^\pm$ by an $SU(2)$ transformation. This $SU(2)$ transformation would act, via its associated $SO(3)$ element, on the complex structures $(\ICS_1,\ICS_2,\ICS_3)$ by rotation. The analogue of the bijection \eqref{bosomaj} defined in terms  of $\tilde{\chi}^\pm$  would then intertwine between the action of the standard complex structures \eqref{hyperaction} on  the bosonic zero-modes and the action of the rotated complex structures on the fermionic zero-modes.


\subsection{The action of the angular momentum operator on zero-modes}
\label{angsect}

All the fields appearing in   $\nsusy=4$ supersymmetric gauge theory are acted on by rotations. For bosonic fields this is an $SO(3)$ action and for fermionic fields it is an $SU(2)$ action. In order to write down the action of finite and infinitesimal rotations on the bosonic and fermionic zero-modes in our quaternionic notation we use the standard relation between unit quaternions on the one hand and $SO(3)$  and $SU(2)$ matrices on the other. We briefly summarise this relation in our notation.

It follows from the quaternion algebra \eqref{quatalg} that the quaternions
\begin{align}
 \qS_j & \ = \ \frac{1}{2}\qq_j
\end{align}
satisfy the  commutation relations
\begin{align}
 [\qS_i,\qS_j] & \ = \ \epsilon_{ijk}\qS_k
\end{align}
of the Lie algebra $su(2)$. It also follows from the algebra \eqref{quatalg} that
\begin{align}
\exp\left(\alpha n_i \qS_i\right) & \ = \ \cos\left(\frac{\alpha}{2}\right) + n_i\qq_i\sin\left( \frac{\alpha}{2}\right),
\end{align}
for any $\alpha \in [0,4\pi)$ and any unit vector $(n_1,n_2,n_3)$. This shows that the exponentiation of linear combinations of the $\qS_i$ gives a unit quaternion, and that any unit quaternion can be written in this way. Inserting the representation \eqref{quatrep}  gives the usual identification of  the set of  unit quaternions with the group $SU(2)$.

Consider now a fixed unit quaternion  $\qu$. The $SO(3)$ matrix $R$ associated to it in the usual way, via the adjoint action, is given by
\begin{align}
 \label{addef}
\qu\qq_i\overline{\qu} & \ = \ R_{ji}\qq_j\eqip{,}
\end{align}
where we used that $\overline{\qu}=\qu^{-1}$ for unit quaternions.
Conversely, a matrix $R\in SO(3)$ determines a pair $\pm \qu$ of unit quaternions via the above relation.  In the following we write $R_\qu$ for the $SO(3)$ matrix associated to a quaternion $\qu$ via \eqref{addef}, and $\qu_R$ for a unit quaternion that satisfies equation~\eqref{addef} for given $R\in SO(3)$ (i.e. we pick one of the two possible choices).

According to the definitions \eqref{defW} and \eqref{bzmq}, a  bosonic zero-mode $\qBos$ involves a scalar field and a vector field on $\RR^3$. The action of a spatial rotation $R\in SO(3)$  on  both of these fields can be summarised in one $SO(3)$ action $\rho^B (R)$ on a  bosonic zero-mode  $\qBos$ in quaternionic notation as
\begin{align}
\label{bosso3}
 \rho^B(R)(\qBos)\, (\vec{x}) & \ \simeq \ \qu_R \, \qBos (R^{-1}(\vec{x})) \, \overline{\qu}_R\eqip{.}
\end{align}
Note that the right-hand side of \eqref{bosso3} does not depend on the sign we choose for $\qu_R$ and  thus only depends on  $R\in SO(3)$.  Here and in the remainder of this section we use  $\simeq$ to indicate an equality up to  gauge transformation (either finite or infinitesimal, which will be clear from the context).
The implementation of symmetries in gauge theories is subtle and typically involves a lifting of the symmetry group into the gauge group. Details are discussed,  for example, in the textbook by Manton and Sutcliffe\mycite{MantonSutcliffe} but do not need to concern us here, since our ultimate goal is a relation between gauge-invariant operators on the moduli space, to be discussed in section~4.

Having defined the action of finite rotations, we obtain the components $J_i$ of the angular momentum operator by computing the effect of infinitesimal rotations around three orthogonal axes.
Writing  $R_i(\eps)$ for the rotation about the $i$-th axis by  $\eps$, so that $R_i(\eps)$  is the $SO(3)$ matrix associated to the quaternion $\exp(\eps \qS_i)$, we define the angular momentum operator via
\begin{align}
 \Jang^{B}_i \qBos & \ = \ i \left.\frac{\d}{\d\eps}\right|_{\eps =0}\rho^B(R_i(\eps))\, (\qBos)\eqip{,} \anumber
\end{align}
which satisfies the angular momentum algebra $\left[ \Jang_i, \Jang_j\right] = i \eps_{ijk} \Jang_k$. Explicitly, we find
\begin{align}
\label{bosonang}
 \Jang^{B}_i \qBos & \ \simeq  \ i \left( {Y_i}\qBos  + [\qS_i,\qBos] \right)\eqip{,}
\end{align}
where $Y_i$ are the vector fields generating rotations of $\RR^3$ about the $i$-th axis.

Next, we turn to the fermionic zero-modes, using their description in terms of Dirac spinors.  Since these are necessarily of the form \eqref{4two2}, we can express the action of rotations on the spinors in terms of the action on the Weyl spinors $\eta^\pm$. Spatial rotations act on these via the usual spinorial $SU(2)$ action
\begin{align}
\rho^F(\qu) (\eta^\pm) \, (\vec{x}) & \ \simeq \ \qu \, \eta^\pm(R_\qu^{-1}(\vec{x}))\eqip{,} \label{fermSU2}
\end{align}
where we have again used the $SO(3)$ matrix $R_\qu$ associated to $\qu$ via \eqref{addef} and we have not notationally distinguished between a quaternion $\qu$ and the $SU(2)$ matrix associated to it via the representation \eqref{quatrep}.

Now we note that we can combine the two Weyl spinors into a complex  $2 \times 2$ matrix $\qBos_\eta=(\eta^+,\eta^-)$, i.e.
\begin{align}
\qBos_\eta & \ = \ \begin{pmatrix} \eta^+_1 &  \eta^-_1 \\ \eta^+_2 &  \eta^-_2\end{pmatrix},
\end{align}
with the components of $\eta^\pm$ denoted $\eta^\pm_1$ and $\eta^\pm_2$.
We can make this explicit by  writing $\qBos_\eta$ in terms of quaternions as
\begin{align}
\label{fermquati}
\qBos_\eta \ =  \ \frac 1 2 \left(\eta^+_1(\qq_4+i\qq_3) + \eta^+_2(\qq_2+i\qq_1)\right)
+\frac 1 2 \left(\eta^-_2(\qq_4-i\qq_3) + \eta^-_1(-\qq_2+i\qq_1)\right).
\end{align}
For later use we note that, with this notation, the action of the complex structures on the fermionic zero-modes \eqref{hyperdirac} can  expressed in terms of $\qBos_\eta$ as for the bosonic zero-modes \eqref{hyperaction}.

The $SU(2)$ action can be written in quaternionic language as
\begin{align}
\rho^F(\qu)(\qBos_\eta)\,(\vec{x}) & \ \simeq \ \qu \qBos_\eta (R_\qu^{-1}(\vec{x})).
\end{align}
The fermionic angular momentum operator is defined via the corresponding infinitesimal action
\begin{align}
 J^{F}_i \qBos_\eta & \ =  \ i \left.\frac{d}{d\eps}\right|_{\eps =0}\rho^F(\exp(\eps\qS_i))(\qBos_\eta)
\end{align}
and therefore given by
\begin{align}
\label{fermionang}
 J^F_i\qBos_\eta & \ \simeq  \ i \left( Y_i\qBos_\eta  + \qS_i\qBos_\eta \right)\eqip{.}
\end{align}

The upshot of the discussion thus far is that we can view both bosonic and fermionic zero-modes as quaternion-valued functions on $\RR^3$ (i.e. as $\qBos$ and $\qBos_{\eta}$ respectively), but that the action of the  angular momentum operator is different in the two cases. Comparing the expressions \eqref{fermionang} and  \eqref{bosonang},  and noting that the action \eqref{hyperdirac} of complex structures on the fermionic zero-modes
can be written in terms of $\qBos_{\eta}$ as
\begin{align}
 \ICS_i:  \qBos_{\eta} & \ \mapsto \ -2 \qBos_{\eta} \qS_i \anumber
\end{align}
we arrive at the fundamental relation
\begin{align}
\label{keyresult}
 \Jang^F_i\qBos_{\eta} & \ \simeq \ \Jang^B_i \qBos_{\eta} - \half[i] \ICS_i \qBos_{\eta}
\end{align}
between the fermionic and bosonic angular momentum operators acting on quaternion-valued functions.

Equation \eqref{keyresult} relates three quantities which all play important roles in this paper. We will discuss this relation in more detail, and in   geometrical language, in section 4  of this paper. At this point, we would like to highlight some basic geometrical aspects.  Since  bosonic zero-modes satisfy the  linearised Bogomol'nyi equations, as in equations \eqref{bosonzero}, they  should be thought of as  (representatives of)  tangent vectors to the moduli space of monopoles, with the linearised Gauss' law ensuring that they are orthogonal to gauge orbits. More precisely, they represent tangent vectors in the tangent space at the point in the moduli space represented by the configuration $W_{\underline{i}}$.
By definition, bosonic zero-modes are required to be normalisable with respect to the natural inner product,
which, in fact,  defines the metric $g$ on the moduli space\footnote{  \emph{cf.} equations (17) and (27) in our previous paper\mycite{deVriesSchroers:susyQMMM}.},
\begin{align}
 \langle \qBos , \qv\rangle & \ = \ g(\qBos, \qv) \ = \ \int \d^3 x \ \Re (\qBos \overline{\qv}), \label{bosInnerProd}
\end{align}
where $\Re$ denotes  the coefficient of $\qq_4$ of the quaternion $\qBos \overline{\qv}$.
Using this inner product, a  bosonic zero-mode $\qBos$ can be viewed as a linear map $  \langle \qBos , \, \cdot \, \rangle $ which maps bosonic zero-modes to real numbers, i.e. as a cotangent vector:
\begin{align}
\label{dualmap}
 \langle \qBos, \, \cdot \, \rangle : \qv & \ \mapsto \ \langle \qBos , \qv\rangle \in \RR \eqip{.}
\end{align}
Viewing bosonic zero-modes as cotangent vectors, the $SO(3)$ action \eqref{bosso3} 
on tangent vectors gives rise to an action on cotangent vectors, which we also denote by $\rho^B$:
\begin{align}
\rho^B(R):  \langle \qBos , \, \cdot \, \rangle & \ \mapsto \ \langle \rho^B(R)(\qBos), \, \cdot \, \rangle\eqip{.} \label{dualbosso3}
\end{align}
The corresponding angular momentum operators, again denoted by $\Jang_i^B$, act via
\begin{align}
\Jang_i^B:   \langle \qBos , \, \cdot \, \rangle   & \ \mapsto \ \langle \Jang_i^B(\qBos), \, \cdot \, \rangle\eqip{.} \label{dualbosonang}
\end{align}

The inner product \eqref{bosInnerProd} can be extended bilinearly to complexified quaternions.
As a result,  the description of fermionic zero-modes in terms of   the complex quaternion-valued  function  $\qBos_{\eta}$ on $\RR^3$ allows us to view fermionic zero-modes as 
maps
\begin{align}
\label{dualmapf}
 \langle \qBos_\eta, \, \cdot \, \rangle : \qv & \ \mapsto \ \langle \qBos_\eta , \qv\rangle \in \CC \eqip{.}
\end{align}
 i.e. as complexified cotangent vectors or 1-forms on the moduli space.
This point of view turns out to be very natural, as we shall see, capturing the fermionic nature of fermionic zero-modes in the anti-commutativity of the wedge-product of 1-forms, and allowing for geometrically natural expressions for the supercharges.
In this context  it is worth noting that  it follows from \eqref{hyperdirac} or \eqref{fermquati}  that  the  form $\langle \qBos_\eta, \, \cdot \, \rangle$
is an anti-holomorphic  form  (eigenvalue $+i$) with respect to $\ICS_3$  when $\eta^- =0$ but $\eta^+\neq0$,
and is a holomorphic form (eigenvalue $-i$)  when $\eta^+ =0$ but $\eta^-\neq0$. As we shall see in section~\ref{sectquantFZM}, this is mirrored in  the quantisation of fermionic zero-modes as  differential forms  on the moduli space.

 When we describe the fermionic zero-modes in terms of differential forms on the moduli space, we have to make sure that we use the fermionic angular momentum operator \eqref{fermionang} and not the geometrically natural (bosonic) operator generating rotations for cotangent vectors \eqref{bosonang}:
\begin{align}
\Jang_i^F:  \langle \qBos_{\eta} , \, \cdot \, \rangle & \ \mapsto \ \langle \Jang_i^F(\qBos_{\eta}), \, \cdot \, \rangle\eqip{.} \label{dualfermionang}
\end{align}
 Comparing  \eqref{dualbosonang} and \eqref{dualfermionang}  we deduce the following relation between the two angular momentum operators acting on forms:
\begin{align}
  \Jang^F_i & \ \simeq \ \Jang^B_i - \half[i] \ICS_i\eqip{,} \label{keyresult2}
\end{align}
with the factor of $i$ in front of $\ICS_i$ in this equation again being  due to our convention  for the angular momentum algebra  $\left[ \Jang_i, \Jang_j\right] = i \eps_{ijk} \Jang_k$.
Thus, expressed in geometrical terms, this relation means that, when viewing fermionic zero-modes as differential 1-forms on the moduli space, we obtain the correct angular momentum operator by subtracting $\half[i] \ICS_i$ from the natural (bosonic) operator generating rotations for cotangent vectors.

\subsection{Vector spaces of zero-modes and their bases}
Denoting the moduli space of charge-$\topk$ monopoles by  $\MS_{\topk}$ we now consider a set of local coordinates $X^a$, $a \in \{1,\ldots, 4k\}$, on $\MS_{\topk}$. The variations $\delta_a W_{\underline i} = \frac{\delta W_{\underline i}}{\delta X^a}$ provide a basis of the vector space of bosonic zero-modes. In quaternionic language this corresponds to the following basis of the bosonic zero-modes:
\begin{align}
\label{bosbas}
\qBos_a & \ = \ \overline{\qq}_{\underline{i}} \, \delta_a W_{\underline{i}}.
\end{align}

The total space of fermionic zero-modes in the $\nsusy=2$ supersymmetric theory is a $2k$-dimensional complex vector space, or a $4k$-dimensional real vector space. In the $\nsusy=4$ supersymmetric theory we have a doubling of the fermionic (Dirac) zero-modes, and hence the space of Dirac zero-modes $V$ is $4k$-dimensional as a complex vector space, or $8k$-dimensional as a real vector space.  Defining  fermionic zero-modes $\xi_a^\pm$  associated to  the  bosonic zero-modes \eqref{bosbas}  as in \eqref{fermbos},  a complex basis for $V$ is given by
\begin{align}
\mathscr{B}_V^{\CC} & \ = \ \{(\xi^+_1,0), \ldots, (\xi^+_{2k},0), \ (0,\xi^-_1), \ldots, (0,\xi^-_{2k}) \}\eqip{.} \label{CbasisV}
\end{align}
Using the notation $\xi^\pm_a$ with $a \in \{1,\ldots, 4k\}$, where $\xi^\pm_a = \ICS(\xi^\pm_{a-2k})$ if $a>2k$, a   real basis for this vector space is given by
\begin{align}
\mathscr{B}_V^{\RR} & \ =  \{(\xi^+_1,0), \ldots, (\xi^+_{4k},0), \ (0,\xi^-_1), \ldots, (0,\xi^-_{4k}) \}\eqip{.} \label{RbasisV}
\end{align}
We shall use this basis in our calculation of the effective Lagrangian below.

As we saw above, each pair of  Dirac zero-modes $(\xi^+,\xi^-)$ corresponds to quartet of Majorana zero-modes $(\psi_1,\psi_2,\psi_3,\psi_4)$.  Since the Majorana zero-modes must satisfy the Majorana condition \eqref{chargecon} (which can be thought of as a reality condition), the resulting basis should again be seen as a basis over the real numbers.
It follows that the  vector space of Majorana zero-modes, like $\mathscr{B}_V^{\RR}$, is an $8k$-dimensional  real vector space.

\subsection{The effective Lagrangian in real coordinates}
In order to compute the effective Lagrangian for the moduli space approximation, we first express the Lagrangian \eqref{LN4}  in terms of the pair of Dirac fields $\xi^\pm$ instead of the quartet of Majorana fermions $\psi_r$.
We approximate time-derivatives of the fields $W_{\underline{i}}$ in terms of the bosonic zero-modes by assuming
\begin{align}
\dot{W}_{\underline{i}} & \ = \ \delta_a W_{\underline{i}} \dot{X}^a\eqip{.} \label{bosonZMN4}
\end{align}
Furthermore, we parameterise  the fermionic zero-modes in terms of the real basis  $\mathscr{B}_V^{\RR} = \{\xi^\pm_a\}$  \eqref{RbasisV} of Dirac fermions. Introducing two  real valued Grassmann numbers $\lambda^a_{\pm}$,
we parameterise the fermionic zero-modes as
\begin{align}
\label{fermZMN4}
\xi^{\pm} & \ = \ \xi^{\pm}_a \lambda_{\pm}^a\eqip{.}
\end{align}
We insert this into the Lagrangian and expand to the lowest non-trivial order, taking Gauss' law into account. The calculations involved in this procedure are well documented in the literature (see Blum\mycite{Blum:SusyQMMonN=4}   and  Gauntlett\mycite{Gauntlett:LowEnDymN=2Mon}, as well as the review paper by Weinberg and Yi\mycite{WeinbergYi:MMonDynSusyDuality} which uses conventions very similar to ours). The result is
\begin{align}
L_{\mbox{eff}} & \ = \ \half g_{ab} \dot{X}^a \dot{X}^b + \half[i] g_{ab} (\lambda^a)^T (D_t \lambda)^b - \frac{1}{8} R_{abcd} (\lambda^a)^T \lambda^b (\lambda^c)^T \lambda^d - \frac{4\pi a}{e} \Bbk\eqip{,} \label{LeffN4}
\end{align}
where the metric is given by $g_{ab} = \int \d^3 x \ \delta_a W_{\underline{i}} \cdot \delta_b W_{\underline{i}}$, and we have combined the parameters $\lambda_+^a$ and $\lambda_-^a$ into the two-component Grassmann function
\begin{align}
  \lambda^a & \ = \ \left(
                      \begin{array}{c}
                        \lambda_+^a \\
                        \lambda_-^a \\
                      \end{array}
                    \right)\eqip{.} \anumber
\end{align}
Compared to the effective Lagrangian of the $\nsusy = 2$ supersymmetric monopole, we now have two copies of the fermionic term in our effective Lagrangian,
\begin{align}
  \half[i] g_{ab} (\lambda^a)^T (D_t \lambda)^b & \ = \ \half[i] g_{ab} \lambda_+^a D_t \lambda_+^b + \half[i] g_{ab} \lambda_-^a D_t \lambda_-^b\eqip{,} \anumber
\end{align}
and an extra term involving the curvature of the metric, with components $R_{abcd}$, which provides a coupling between the fermionic variables $\lambda_+^a$ and $\lambda_-^a$.

\section{Quantisation of the effective Lagrangian}
\label{sectN4FZMFMS}

As explained in the introduction, we would like to adopt a quantisation procedure that brings out, as clearly as possible, the doubling of fermionic zero-modes from the $\nsusy=2$ to the $\nsusy =4$ theory.
For this purpose, it is best to write the effective Lagrangian in terms of complex coordinates. Using complex coordinates on the moduli space, the two sets of fermionic zero-modes will be clearly distinguishable: one of them we shall identify as holomorphic forms, and the other we shall identify as anti-holomorphic forms. This is different from the usual quantisation procedure (which uses real coordinates on the moduli space, see for example Weinberg and Yi\mycite{WeinbergYi:MMonDynSusyDuality}) where this distinction is not apparent.

\subsection{Complex coordinates and an orthogonal frame}
We choose $2k$ (local) complex coordinates $Z^{\alpha}$ for the hyperk\"ahler manifold $\MS_{\topk}$ which are holomorphic with respect to the complex structure $\ICS = \ICS_3$. The real and imaginary parts of $Z^{\alpha}$ form a  set of real coordinates $X^a$ (the index $a$, as usual, runs from $1$ to $4k$).
We may choose these coordinates so that, locally,
\begin{align}
  Z^{\alpha} \ = \ & X^{\alpha} + i X^{\alpha + 2k}\eqip{,} &
  \overline{Z}^{\overline{\alpha}} \ = \ & X^{\alpha} - i X^{\alpha + 2k}\eqip{,} \anumber
\end{align}
where the index $\alpha$ runs from $1$ to $2k$. We denote the metric in complex coordinates by $\sfg$. Since the moduli space is Hermitian, its components satisfy $\sfg_{\alpha \beta} = \sfg_{\overline{\alpha}\overline{\beta}} = 0$.

It is also convenient to work with an orthonormal frame to parameterise the fermionic zero-modes\mycite{Gauntlett:LowEnDymSusySol,deVriesSchroers:susyQMMM}. We introduce an orthonormal frame $\ofc$, choosing it so that it respects holomorphicity,
\begin{align}
\sfg_{\alpha\overline{\beta}} & \ = \ \delta_{A\overline{B}} {\ofc^A}_{\alpha} {\ofc^{\overline{B}}}_{\overline{\beta}}\eqip{,} &
\ofc^A & \ = \ {\ofc^A}_{\alpha} \d Z^{\alpha}\eqip{,} &
\ofc^{\overline{A}} & \ = \ {\ofc^{\overline{A}}}_{\overline{\alpha}} \d \overline{Z}^{\overline{\alpha}}\eqip{,} \anumber
\end{align}
where the indices $A$ and $\overline{B}$ run from $1$ to $2k$, $\delta_{A\overline{B}} = 1$ if $A = B$, and $\delta_{A\overline{B}} = 0$ if $A \neq B$. We now define
\begin{align}
  \zeta^{A}_{\pm} \ = \ & \lambda^{A}_{\pm} + i \lambda^{A + 2k}_{\pm}\eqip{,} &
  \zeta^{\overline{A}}_{\pm} \ = \ & \lambda^{A}_{\pm} - i \lambda^{A + 2k}_{\pm}\eqip{,} \label{defzetapm}
\end{align}
and write the effective Lagrangian \eqref{LeffN4} in terms of these variables. Using the tensorial nature of each term, the coordinate transformations take on a simple form.
Since $\sfg_{\alpha \beta} = \sfg_{\overline{\alpha}\overline{\beta}} = 0$, we find for the first term $\half g_{ab} \dot{Z}^{a} \dot{Z}^{b} = \half (\sfg_{\alpha \overline{\beta}} \dot{Z}^{\alpha} \dot{\overline{Z}}^{\overline{\beta}} + \sfg_{\overline{\alpha} \beta} \dot{\overline{Z}}^{\overline{\alpha}} \dot{Z}^{\beta})$, and we may use the symmetry of the metric, $\sfg_{\alpha \overline{\beta}} = \sfg_{\overline{\beta} \alpha}$, to simplify the outcome.
After transforming the second term, we also use partial integration of the time-derivative to simplify the resulting expression.
For the curvature term, we use the fact that most components of the curvature tensor on a K\"ahler manifold vanish\mycite{Moroianu}. The non-vanishing components of the curvature tensor on a K\"ahler manifold are $R_{\alpha\overline{\beta}\gamma\overline{\delta}}$, $R_{\overline{\alpha}\beta\gamma\overline{\delta}}$, $R_{\alpha\overline{\beta}\overline{\gamma}\delta}$ and $R_{\overline{\alpha}\beta\overline{\gamma}\delta}$ and they have the following symmetries:
\begin{align}
R_{\alpha\overline{\beta}\gamma\overline{\delta}} &
\ = \ - R_{\overline{\beta}\alpha\gamma\overline{\delta}}
\ = \ - R_{\alpha\overline{\beta}\overline{\delta}\gamma}
\ = \ R_{\overline{\beta}\alpha\overline{\delta}\gamma}\eqip{,} &
R_{\alpha\overline{\beta}\gamma\overline{\delta}} &
\ = \ R_{\gamma\overline{\delta}\alpha\overline{\beta}}\eqip{.} \anumber
\end{align}
Altogether, the Lagrangian becomes
\begin{align}
L_{\mbox{eff}} & \ = \ \sfg_{\alpha \overline{\beta}} \dot{Z}^{\alpha} \dot{\overline{Z}}^{\overline{\beta}} + i \delta_{\overline{A} B} (\zeta^{\overline{A}})^T \left( D_t \zeta \right)^{B} - \half R_{A\overline{B}C \overline{D}} (\zeta^A)^T \zeta^{\overline{B}} (\zeta^C)^T \zeta^{\overline{D}} - \frac{4\pi a}{e}\Bbk\eqip{.} \anumber
\end{align}
The canonical momenta are therefore given by
\begin{align}
P_{\alpha} \ = \ \frac{\partial L_{\mbox{eff}}}{\partial \dot{Z}^{\alpha}} & \ = \ \sfg_{\alpha \overline{\beta}} \dot{\overline{Z}}^{\overline{\beta}} + i \omega_{\alpha\overline{A}C} (\zeta^{\overline{A}})^T \zeta^C\eqip{,} \nonumber \\
P_{\overline{\alpha}} \ = \ \frac{\partial L_{\mbox{eff}}}{\partial \dot{\overline{Z}}^{\overline{\alpha}}} & \ = \ \sfg_{\beta \overline{\alpha}} \dot{Z}^{\beta} + i \omega_{\overline{\alpha}\overline{A}C} (\zeta^{\overline{A}})^T \zeta^{C}\eqip{,} \nonumber \\
\Pi^{\pm}_{A} \ = \ \frac{\partial L_{\mbox{eff}}}{\partial \dot{\zeta}_{\pm}^{A}} & \ = \ - i \delta_{\overline{B}A} \zeta_{\pm}^{\overline{B}}\eqip{,} \nonumber \\
\Pi^{\pm}_{\overline{A}} \ = \ \frac{\partial L_{\mbox{eff}}}{\partial \dot{\zeta}_{\pm}^{\overline{A}}} & \ = \ 0\eqip{.} \anumber
\end{align}
The effective Hamiltonian is then
\begin{align}
H_{\mbox{eff}} & \ = \ \dot{Z}^{\alpha} P_{\alpha} + \dot{Z}^{\overline{\alpha}} P_{\overline{\alpha}} + (\dot{\zeta}^A)^T \Pi_A + (\dot{\zeta}^{\overline{A}})^T \Pi_{\overline{A}} - L_{\mbox{eff}} \ = \ H_0 + \frac{4\pi a}{e} \Bbk\eqip{,} \anumber
\end{align}
where we have defined
\begin{align}
H_0 & \ = \ \half \sfg^{\alpha\overline{\beta}} \tilde{P}_{\alpha} \tilde{P}_{\overline{\beta}} + \half
 R_{ A \overline{B} C \overline{D} } (\zeta^A)^T \zeta^{\overline{B}} (\zeta^C)^T \zeta^{\overline{D}}\eqip{,} \anumber
\end{align}
and
\begin{align}
  \tilde{P}_{\alpha} & \ = \ P_{\alpha} - i \omega_{\alpha\overline{A}C} (\zeta^{\overline{A}})^T \zeta^C \ = \ \sfg_{\alpha\overline{\beta}} \dot{\overline{Z}}^{\overline{\beta}}\eqip{,} \nonumber \\
  \tilde{P}_{\overline{\alpha}} & \ = \ P_{\overline{\alpha}} - i \omega_{\overline{\alpha}\overline{A}C} (\zeta^{\overline{A}})^T \zeta^{C} \ = \ \sfg_{\overline{\alpha}\beta} \dot{Z}^{\beta}\eqip{.} \anumber
\end{align}

\subsection{Quantisation: fermionic zero-modes as forms on the moduli space}
\label{sectquantFZM}
The expressions for the fermionic momenta above lead to constraints in the Poisson algebra. As a result, we need to replace Poisson brackets with Dirac brackets before we can quantise. To quantise the canonical coordinates and the supersymmetry charges we shall take this route. The analysis in this section closely follows the corresponding analysis of the $\nsusy=2$ case in our previous paper\mycite{deVriesSchroers:susyQMMM}.

The only non-vanishing Dirac brackets between the canonical coordinates are
\begin{align}
\left\{ P_{\alpha} , Z^{\beta} \right\}_{DB} & \ = \ \delta_{\alpha}^{\beta}\eqip{,} &
\left\{ P_{\overline{\alpha}} , \overline{Z}^{\overline{\beta}} \right\}_{DB} & \ = \ \delta_{\overline{\alpha}}^{\overline{\beta}}\eqip{,} \nonumber  \\
\left\{ \zeta_{+}^A , \zeta_{+}^{\overline{B}} \right\}_{DB} & \ = \ i \delta^{A\overline{B}}\eqip{,} &
\left\{ \zeta_{-}^A , \zeta_{-}^{\overline{B}} \right\}_{DB} & \ = \ i \delta^{A\overline{B}}\eqip{.} \anumber
\end{align}
The bosonic brackets are precisely the same as in the $\nsusy=2$ case, and fermionic ones simply consist of two copies (labelled $+$ and $-$) of the $\nsusy=2$ case. We can therefore quantise the Dirac brackets in the same way as in the $\nsusy=2$ case:
\begin{align}
\left\{ P_{\alpha} , Z^{\beta} \right\}_{DB} & \ = \ \delta_{\alpha}^{\beta} & \mapsto && \left[ P_{\alpha} , Z^{\beta} \right] & \ = \ -i \delta_{\alpha}^{\beta} \nonumber \\
\left\{ P_{\overline{\alpha}} , \overline{Z}^{\overline{\beta}} \right\}_{DB} & \ = \ \delta_{\overline{\alpha}}^{\overline{\beta}} & \mapsto && \left[ P_{\overline{\alpha}} , \overline{Z}^{\overline{\beta}} \right] & \ = \ - i \delta_{\overline{\alpha}}^{\overline{\beta}} \nonumber \\
\left\{ \zeta^A_{+} , \zeta^{\overline{B}}_{+} \right\}_{DB} & \ = \ i \delta^{A\overline{B}} & \mapsto && \left\{ \zeta^A_{+} , \zeta^{\overline{B}}_{+} \right\} & \ = \ \delta^{A\overline{B}} \nonumber \\
\left\{ \zeta^A_{-} , \zeta^{\overline{B}}_{-} \right\}_{DB} & \ = \ i \delta^{A\overline{B}} & \mapsto && \left\{ \zeta^A_{-} , \zeta^{\overline{B}}_{-} \right\} & \ = \ \delta^{A\overline{B}} \anumber
\end{align}
In the $\nsusy=2$ case, a representation of the algebra spanned by the momenta, coordinates and fermionic coordinates was formed using the space of anti-holomorphic forms on the moduli space of monopoles. To accommodate the two sets of the fermionic zero-modes in the $\nsusy=4$ case, we now need to include holomorphic forms as well. (This assumes we have made the appropriate choice for $\chi^+$ and $\chi^-$ in equations \eqref{choicechiN4} to guarantee the right behaviour of fermionic zero-modes under the complex structures; see also the discussion at the end of section~\ref{sectComplStructAct}.) Therefore we take the Hilbert space of states to be the space of all square-integrable differential forms on the moduli space, including both holomorphic and anti-holomorphic forms.

The bosonic coordinates act by multiplication and the bosonic momenta are represented as derivatives,
\begin{align}
P_{\alpha} & \ \mapsto \ -i \partial_{\alpha} &
P_{\overline{\alpha}} & \ \mapsto \ - i \partial_{\overline{\alpha}} \anumber
\end{align}
while the quantisation of fermions is given by
\begin{align}
\label{fermquant}
\zeta^{\overline{A}}_{+} & \ \mapsto \ \ofc^{\overline{A}} \wedge &
\zeta^{A}_{+} & \ \mapsto \ \iota(\ofc^{A}) \nonumber \\
\zeta^{A}_{-} & \ \mapsto \ \ofc^{A} \wedge &
\zeta^{\overline{A}}_{-} & \ \mapsto \ \iota(\ofc^{\overline{A}})
\end{align}
where $\iota(\ofc^{A})(\ofc^{\overline{B}}) = \delta^{A \overline{B}}$, $\iota(\ofc^{\overline{A}})(\ofc^{B}) = \delta^{\overline{A} B}$ and $\iota(\ofc^{A})(\ofc^{B}) = \iota(\ofc^{\overline{A}})(\ofc^{\overline{B}}) = 0$.

The covariant momenta are quantised, as in the $\nsusy=2$ case, as covariant derivatives:
\begin{align}
  \tilde{P}_{\alpha} & \ = \ P_{\alpha} - i \omega_{\alpha\overline{A}C} (\zeta^{\overline{A}})^T \zeta^C & \mapsto && -i \nabla_{\alpha}\eqip{,} \nonumber \\
  \tilde{P}_{\overline{\alpha}} & \ = \ P_{\overline{\alpha}} - i \omega_{\overline{\alpha}\overline{A}C} (\zeta^{\overline{A}})^T \zeta^{C} & \mapsto && -i \nabla_{\overline{\alpha}}\eqip{.}
\end{align}

\subsection{Supersymmetry}
\label{sectN4FZMSSusy}
 The effective action corresponding to the effective Lagrangian \eqref{LeffN4} is invariant under $\nsusyms = 8$ supersymmetry transformations\mycite{WeinbergYi:MMonDynSusyDuality}.
Writing these in terms of complex coordinates, we find
\label{susyTransCompN2}\begin{align}
\delta_{\unit} Z^{\alpha} & \ = \ \phantom{-i} \overline{\eps} \zeta^{\alpha} &
\delta_{\unit} \zeta^{\alpha} & \ = \ \phantom{-} i \dot{Z}^{\alpha} \ \pauli_2 \eps \nonumber \\
\delta_{\unit} \overline{Z}^{\overline{\alpha}} & \ = \ \phantom{-i} \overline{\eps} \zeta^{\overline{\alpha}} &
\delta_{\unit} \zeta^{\overline{\alpha}} & \ = \ \phantom{-} i \dot{\overline{Z}}^{\overline{\alpha}} \ \pauli_2 \eps \nonumber 
\\
\delta_\ICS Z^{\alpha} & \ = \ \phantom{-}i \overline{\eps} \zeta^{\alpha} &
\delta_\ICS \zeta^{\alpha} & \ = \ \phantom{-i} \dot{Z}^{\alpha} \ \pauli_2 \eps \nonumber \\
\delta_\ICS \overline{Z}^{\overline{\alpha}} & \ = \ -i \overline{\eps} \zeta^{\overline{\alpha}} &
\delta_\ICS \zeta^{\overline{\alpha}} & \ = \ -\phantom{i} \dot{\overline{Z}}^{\overline{\alpha}} \ \pauli_2 \eps \nonumber 
 \\
\delta_\JCS Z^{\alpha} & \ = \ \overline{\eps} {\JCS^{\alpha}}_{\overline{\beta}} \zeta^{\overline{\beta}} &
\delta_\JCS \zeta^{\alpha} & \ = \ -i {\JCS^{\alpha}}_{\overline{\beta}} \dot{\overline{Z}}^{\overline{\beta}} \ \pauli_2 \eps - \Gamma^{\alpha}_{\beta\gamma} \left( \overline{\eps} {\JCS^{\beta}}_{\overline{\delta}} \zeta^{\overline{\delta}} \right) \zeta^{\gamma} \nonumber \\
\delta_\JCS \overline{Z}^{\overline{\alpha}} & \ = \ \overline{\eps} {\JCS^{\overline{\alpha}}}_{\beta} \zeta^{\beta} &
\delta_\JCS \zeta^{\overline{\alpha}} & \ = \ -i {\JCS^{\overline{\alpha}}}_{\beta} \dot{Z}^{\beta} \ \pauli_2 \eps - \Gamma^{\overline{\alpha}}_{\overline{\beta}\overline{\gamma}} \left( \overline{\eps} {\JCS^{\overline{\beta}}}_{\delta} \zeta^{\delta} \right) \zeta^{\overline{\gamma}} \nonumber 
\\
\delta_\KCS Z^{\alpha} & \ = \ \overline{\eps} {\KCS^{\alpha}}_{\overline{\beta}} \zeta^{\overline{\beta}} &
\delta_\KCS \zeta^{\alpha} & \ = \ -i {\KCS^{\alpha}}_{\overline{\beta}} \dot{\overline{Z}}^{\overline{\beta}} \ \pauli_2 \eps - \Gamma^{\alpha}_{\beta\gamma} \left( \overline{\eps} {\KCS^{\beta}}_{\overline{\delta}} \zeta^{\overline{\delta}} \right) \zeta^{\gamma} \nonumber \\
\delta_\KCS \overline{Z}^{\overline{\alpha}} & \ = \ \overline{\eps} {\KCS^{\overline{\alpha}}}_{\beta} \zeta^{\beta} &
\delta_\KCS \zeta^{\overline{\alpha}} & \ = \ -i {\KCS^{\overline{\alpha}}}_{\beta} \dot{Z}^{\beta} \ \pauli_2 \eps - \Gamma^{\overline{\alpha}}_{\overline{\beta}\overline{\gamma}} \left( \overline{\eps} {\KCS^{\overline{\beta}}}_{\delta} \zeta^{\delta} \right) \zeta^{\overline{\gamma}} \label{susyTransCompN2d}
\end{align}
where $\eps$ are two-component Grassmann parameters and $\overline{\eps} = \eps^T \pauli_2$. The corresponding two-component supercharges are
\begin{align}
\QC_{\unit} & \ = \ \phantom{i} \tilde{P}_{\alpha} \zeta^{\alpha} + \phantom{i} \tilde{P}_{\overline{\alpha}} \zeta^{\overline{\alpha}}\eqip{,} &
\QC_\JCS & \ = \ \tilde{P}_{\overline{\alpha}} {\JCS^{\overline{\alpha}}}_{\alpha} \zeta^{\alpha} + \tilde{P}_{\alpha} {\JCS^{\alpha}}_{\overline{\alpha}} \zeta^{\overline{\alpha}}\eqip{,} \nonumber \\
\QC_\ICS & \ = \ i \tilde{P}_{\alpha} \zeta^{\alpha} - i \tilde{P}_{\overline{\alpha}} \zeta^{\overline{\alpha}}\eqip{,} &
\QC_\KCS & \ = \ \tilde{P}_{\overline{\alpha}} {\KCS^{\overline{\alpha}}}_{\alpha} \zeta^{\alpha} + \tilde{P}_{\alpha} {\KCS^{\alpha}}_{\overline{\alpha}} \zeta^{\overline{\alpha}}\eqip{,} \anumber
\end{align}
which generate the supersymmetry transformations via Dirac brackets\footnote{This is in fact the route we used to derive the supersymmetry transformations in complex coordinates. We first rewrote the supercharges given by Weinberg and Yi\mycite{WeinbergYi:MMonDynSusyDuality} in complex coordinates and proceeded to derive the supersymmetry transformations \eqref{susyTransCompN2d} using Dirac brackets.}. They satisfy the following algebra:
\begin{align}
\left\{ Q^+_{\unit}, Q^+_{\unit} \right\}_{DB} & \ = \ \left\{ Q^-_{\unit}, Q^-_{\unit} \right\}_{DB} \ = \ 2i H_0 \eqip{,} \nonumber \\
\left\{ Q^+_{\ICS_i}, Q^+_{\ICS_j} \right\}_{DB} & \ = \ \left\{ Q^-_{\ICS_i}, Q^-_{\ICS_j} \right\}_{DB} \ = \ \delta_{ij} \ 2i H_0\eqip{,} \label{DBSCN4}
\end{align}
and all other brackets vanishing.

For each component of the supercharges we define complex linear combinations, analogous to those in the $\nsusy=2$ supersymmetric case, by
\begin{align}
\label{superchargesN4}
\tilde{\QC}^{\pm} & \ = \ \phantom{-}\frac{i}{2} (Q^{\pm}_{\unit} + i Q^{\pm}_\ICS) \ = \ \phantom{-}i \tilde{P}_{\overline{\alpha}} \zeta^{\overline{\alpha}}_{\pm} \eqip{,} \nonumber \displaybreak[2] \\
(\tilde{\QC}^{\pm})^* & \ = \ -\frac{i}{2} (Q^{\pm}_{\unit} - i Q^{\pm}_{\ICS}) \ = \ -i \tilde{P}_\alpha \zeta^{\alpha}_{\pm}\eqip{,} \displaybreak[2] \nonumber \\
\tilde{\QC}_\JCS^{\pm} & \ = \ \phantom{-}\frac{i}{2} (Q^{\pm}_\JCS - i Q^{\pm}_\KCS) \ = \ \phantom{-}i \tilde{P}_{\alpha} {\JCS^{\alpha}}_{\overline{\alpha}} \zeta^{\overline{\alpha}}_{\pm}\eqip{,} \nonumber \\
(\tilde{\QC}_\JCS^{\pm})^* & \ = \ -\frac{i}{2} (Q^{\pm}_\JCS + i Q^{\pm}_\KCS) \ = \ -i \tilde{P}_{\overline{\alpha}} {\JCS^{\overline{\alpha}}}_{\alpha} \zeta^{\alpha}_{\pm}\eqip{.}
\end{align}
From \eqref{DBSCN4} we find that the algebra they satisfy, has the following non-vanishing Dirac brackets:
\begin{align}
  \left\{ \tilde{\QC}^{\pm} , (\tilde{\QC}^{\pm})^* \right\}_{DB} & \ = \ i H_0\eqip{,} &
  \left\{ \tilde{\QC}_{\JCS}^{\pm} , (\tilde{\QC}_{\JCS}^{\pm})^* \right\}_{DB} & \ = \ i H_0\eqip{,} \anumber
\end{align}
Building on earlier work by Gauntlett\mycite{Gauntlett:LowEnDymN=2Mon}, we showed in our previous paper\mycite{deVriesSchroers:susyQMMM} that  the quantised supercharges of the $\nsusy=2$ theory are given by (twisted) Dolbeault operators. Here we find corresponding expressions for the quantisation of the additional charges in the $\nsusy=4$ theory:
\begin{align}
\label{twistdouN4}
\tilde{\QC}^+ & \ = \ \phantom{-}i \tilde{P}_{\overline{\alpha}} \zeta_+^{\overline{\alpha}} & \mapsto && \ofc^{\overline{\alpha}} \wedge \nabla_{\overline{\alpha}} & \ = \ \overline{\partial}\eqip{,} \nonumber \\
  (\tilde{\QC}^+)^* & \ = \ -i \tilde{P}_\alpha \zeta_+^{\alpha} & \mapsto && - \iota(\ofc^{\alpha}) \nabla_{\alpha} & \ = \ \overline{\partial}^{\dag}\eqip{,} \nonumber \\
\tilde{\QC}_\JCS^+ & \ = \ \phantom{-}i \tilde{P}_{\alpha} {\JCS^{\alpha}}_{\overline{\alpha}} \zeta^{\overline{\alpha}}_+  & \mapsto && \JCS(\ofc^{\alpha}) \ \wedge \nabla_{\alpha} & \ = \ \overline{\partial}_{\JCS}\eqip{,} \nonumber\\
  (\tilde{\QC}_\JCS^+)^* & \ = \ -i \tilde{P}_{\overline{\alpha}} {\JCS^{\overline{\alpha}}}_{\alpha} \zeta^{\alpha}_+ & \mapsto && - \iota\left(\JCS (\ofc^{\overline{\alpha}})\right)  \nabla_{\overline{\alpha}} & \ = \ \overline{\partial}_{\JCS}^{\dag}\eqip{,}
\end{align}
and
\begin{align}
\label{twistdou}
(\tilde{\QC}^-)^* & \ = \ \phantom{-}i \tilde{P}_{\alpha} \zeta_-^{\alpha} & \mapsto && \ofc^{\alpha} \wedge \nabla_{\alpha} & \ = \ \partial\eqip{,} \nonumber \\
  \tilde{\QC}^- & \ = \ -i \tilde{P}_{\overline{\alpha}} \zeta_-^{\overline{\alpha}} & \mapsto && - \iota(\ofc^{\overline{\alpha}}) \nabla_{\overline{\alpha}} & \ = \ \partial^{\dag}\eqip{,} \nonumber \\
(\tilde{\QC}_\JCS^-)^* & \ = \ \phantom{-}i \tilde{P}_{\overline{\alpha}} {\JCS^{\overline{\alpha}}}_{\alpha} \zeta^{\alpha}_-  & \mapsto && \JCS(\ofc^{\overline{\alpha}}) \ \wedge \nabla_{\overline{\alpha}} & \ = \ \partial_{\JCS}\eqip{,}  \nonumber \\
  \tilde{\QC}_\JCS^- & \ = \ -i \tilde{P}_{\alpha} {\JCS^{\alpha}}_{\overline{\alpha}} \zeta^{\overline{\alpha}}_- & \mapsto && - \iota\left(\JCS (\ofc^{\alpha})\right)  \nabla_{\alpha} & \ = \ \partial_{\JCS}^{\dag}\eqip{.} \anumber
\end{align}
The twisted Dolbeault operators $\partial_{\JCS}$ and $\overline{\partial}_{\JCS}$ and their adjoints are explicitly given  by
\begin{align}
  \partial_{\JCS} & \ = \ \JCS \overline{\partial} \JCS^{-1}\eqip{,} &
  \partial_{\JCS}^{\dag} & \ = \ \JCS \overline{\partial}^{\dag} \JCS^{-1}\eqip{,} \nonumber \\
  \overline{\partial}_{\JCS} & \ = \ \JCS \partial \JCS^{-1}\eqip{,} &
  \overline{\partial}_{\JCS}^{\dag} & \ = \ \JCS \partial^{\dag} \JCS^{-1}\eqip{.} \label{defTwistDolb}
\end{align}
In this way we  arrive at expressions for  all of the  supercharges  in terms of  (twisted) Dolbeault operators. Finally, quantising the Hamiltonian we find again the Laplace operator:
\begin{align}
  H_0 && \mapsto && \left\{ \tilde{\QC}^{\pm} , (\tilde{\QC}^{\pm})^* \right\} \ = \ \overline{\partial} \, \overline{\partial}^{\dag} + \overline{\partial}^{\dag} \overline{\partial}
  \ = \ \partial \, \partial^{\dag} + \partial^{\dag} \partial
  \ = \ \half \Delta\eqip{.} \anumber
\end{align}

%

\subsection{An interim summary}
The hyperk\"ahler structure of the Euclidean 4-space $\RR^4$ provides the field-theoretical origin of hyperk\"ahler structure of the monopole moduli space. It leads to the action of the complex structures on the bosonic and fermionic zero-modes of the field theory, explicitly given in equation \eqref{hyperaction} and equations \eqref{hypermaj} and \eqref{hyperdirac} respectively. We have used the letters $\ICS$, $\JCS$ and $\KCS$, introduced in equations \eqref{rename}, to denote the action of the complex structures on the zero-modes for the complex structures on the moduli space as well. These complex structures act on the space of differential forms, with anti-holomorphic forms being eigenstates of $\ICS$ with eigenvalue $i$,  and holomorphic forms being eigenstates of $\ICS$ with eigenvalue $-i$. This is consistent with the behaviour of the fermionic zero-modes $\xi^\pm$ given in equation \eqref{hyperdirac} and our quantisation of the fermionic coordinates in terms of differential forms \eqref{fermquant}.

The resulting picture is natural from a geometric point of view and can be summarised as follows. The moduli space approximation of the $\nsusy = 4$ supersymmetric Lagrangian in (3+1)-dimensions leads to an $\nsusyms = 8$ supersymmetric $\sigma$-model on the moduli space. Quantum states in the effective model can be interpreted as differential forms on the moduli space. The supercharges corresponding to the $\nsusyms = 8$ supersymmetries are the Dolbeault operator and the $\JCS$-twisted Dolbeault operator, their complex conjugates, and the adjoints of all these operators. The effective Hamiltonian in this geometrical interpretation is half the Laplacian acting on these differential forms. All forms and operators (states, supercharges and the Hamiltonian) are acted upon by the complex structures in a natural way.


\section{Angular momentum and the spin operator}
\label{sectAngMom}

In this section  we translate our results  from section~\ref{angsect} about the total angular momentum operator into geometrical language and derive an expression for the angular momentum operator as a differential operator on  the  moduli space.  We also address the issue of splitting the total angular momentum  into orbital  and spin contributions.
For general magnetic monopoles  one cannot unambiguously separate orbital angular momentum from spin due to the extended nature of monopoles. However, in special cases, in particular for the charge-1 monopole and for well-separated monopoles,  we expect the spin operator to be well-defined.

Osborn\mycite{Osborn:TopChSpin} derived the following expression for
the spin operator for the fermionic zero-modes of the $\nsusy = 4$ supersymmetric $SU(2)$ monopole of charge 1 in terms of fermionic creation and annihilation operators,
\begin{align}
\vec{S} & \ = \ \half \sum_{n,s,s'} {a^n_s}^{\dag} (\vec{\sigma})_{ss'} a^n_{s'}\eqip{,} \label{spinOpFT}
\end{align}
where $n$ labels the fermion species ($n \in \{1,2\}$), while $s$ and $s'$ label the spin states of the fermion zero-mode (i.e. up or down). The zero-modes are modelled by the states generated by the ${a^n_s}^{\dag}$ acting on a vacuum state \ket{0}, defined by $a^n_s \ket{0} = 0$. The resulting states can be grouped into five singlets, four doublets and a triplet under the action of the spin operator.
We will  propose  an expression for  a spin-operator for the charge-1 monopole from our  general  expression for the angular momentum operator in subsection~\ref{sectEx1Spin}, and confirm that it agrees with Osborn's spin operator \eqref{spinOpFT}.

\subsection{The total angular momentum operator on the moduli space}
\label{sectAngMomOp}

The monopole moduli spaces  parameterise gauge equivalence classes of  bosonic minimal-energy field configurations and thus inherit an $SO(3)$ action from the spatial rotations  acting on the bosonic fields of the field theory. This $SO(3)$ action plays an important role in the study of the Riemannian geometry of moduli spaces\mycite{AtiyahHitchin}: it  preserves the metric on the moduli spaces, rotates the complex structures into each other (we give details below), and naturally acts on differential forms and differential operators defined on the moduli spaces, including the Laplace operator and (twisted) Dolbeault operators.

In the field theory, the infinitesimal version of the  $SO(3)$ action on bosonic fields leads to the bosonic angular momentum operator \eqref{bosonang} on bosonic zero-modes. On a given moduli space, the infinitesimal  action of the $SO(3)$ generators defines
vector fields  $\sogen_i$  which we can normalise  so that  $[\sogen_i, \sogen_j] = \epsilon_{ijk} \sogen_k$. The associated Lie derivatives    $\mathcal{L}_{\sogen_i}$  implement the infinitesimal $SO(3)$ action on all geometrical objects naturally associated to the moduli space. In particular, their action on tangent vector fields and 1-forms represent the bosonic angular momentum operators \eqref{bosonang} and \eqref{dualbosonang} discussed in section~\ref{angsect}. Adapting the  notational conventions of that
section,  we thus have the following expression for the  induced bosonic angular momentum operator on the moduli space:
\begin{align}
\label{bosangmod}
J^{\mathcal{B}}_i= i \mathcal{L}_{\sogen_i }.
\end{align}

 As discussed in section~\ref{angsect},  the  natural $SO(3)$ action on 1-forms (or cotangent vectors) does not  correctly describe the transformation behaviour of fermionic zero-modes under rotations.
Since we use differential forms on the moduli space to describe fermionic zero-modes, we need to modify the bosonic operator $J^{\mathcal{B}}_i$ in order  to obtain the physically correct angular momentum operator for fermions. According to the  formula \eqref{keyresult2}  we should thus use the expression
\begin{align}
\label{fermangmod}
J^{\mathcal{F}}_i & \ = \ i  \mathcal{L}_{\sogen_i} - \frac{i}{2} \ICS_i \eqip{}
\end{align}
for the angular momentum operator acting on 1-forms on the moduli space  which respresent fermionic zero modes. Extending this operator to  tensor products of fermionic zero-modes according  to the Leibniz rule is equivalent to  replacing the complex structures $\ICS_i$ by their
the adjoint action\mycite{Verbitsky:HypHolBundHKM} on p-forms.
 We thus   arrive at the following  expression for the angular momentum operator $\vec{\Jang}$  acting on forms and functions on the moduli space, which unifies and extends the bosonic \eqref{bosangmod} and fermionic expressions \eqref{fermangmod}:
\begin{align}
\Jang_i & \ = \ i \left( \mathcal{L}_{\sogen_i} - \frac{1}{2} \ad \ICS_i\right) \eqip{.} \label{Jang}
\end{align}
This formula for the angular momentum operator on the moduli space agrees with the expression proposed in a particular case, and in different notation,  by Bak, Lee and Yi\mycite{BakLeeYi:quantumdyons}. It is also precisely the angular momentum operator which we proposed in  our study\mycite{deVriesSchroers:susyQMMM} of the  $\nsusy = 2$ case (where we had not derived it from the field theory).

Using the fact that the $SO(3)$ action on the moduli space rotates the complex structures, i.e.
$[\mathcal{L}_{\sogen_i},\ICS_{j}]=\epsilon_{ijk}\ICS_{k}$, a straightforward calculation shows that $[\Jang_i, \ad \ICS_j] = 0$.
It follows that the angular momentum operator defined by equation \eqref{Jang} obeys the required angular momentum algebra
\begin{align}
\left[ \Jang_i, \Jang_j \right] & \ = \ i \, \epsilon_{ijk} \Jang_k\eqip{.} \label{angMomAlgebra}
\end{align}
 It also follows that the three generators $\Jang_i$ map (anti-) holomorphic forms to (anti-) holomorphic forms with respect to $\ICS_3$ (even though $\Jang_1$ and $\Jang_2$ are made up of two operators which, individually, map anti-holomorphic forms to holomorphic forms and vice versa). This was a key motivation for proposing definition \eqref{Jang} in the context of the $\nsusy=2$ model, where the Hilbert space contains only anti-holomorphic forms. We will now show that our angular momentum operator  has  the correct commutation relations with the supercharges in the $\nsusy=4$ theory as well.

\subsection{Supercharges and the angular momentum operator}
\label{sectSupChAndJ}
The supercharges can be used to create spin states from bosonic states, and therefore they correspond to spin-$\half$ operators. As such, the angular momentum operators must obey the appropriate algebra with the supercharges.
To derive the commutators of the $\Jang_i$ with the supercharges, i.e. the (twisted) Dolbeault operators, we follow the same approach as in the $\nsusy=2$ case\mycite{deVriesSchroers:susyQMMM}. We first write the (twisted) Dolbeault operators, defined in equations \eqref{defTwistDolb},  in terms of twisted exterior derivatives\mycite{Verbitsky:HypHolBundHKM, Verbitsky:QDolbCompl}
\begin{align}
  \partial & \ = \ \half ( \d + i \d_\ICS )\eqip{,} &
  \partial_{\JCS} & \ = \ \half ( \d_\JCS + i \d_\KCS )\eqip{,} \label{DolbeaultExtDer}
\end{align}
where the twisted exterior derivative\footnote{Viewing the moduli space as a K\"ahler manifold with complex structure $\ICS$, we have $d_\ICS = \ICS \d \ICS^{-1} = d^c$ in the standard notation.} is defined by $\d_{\ICS_i} = {\ICS_i} \d {\ICS_i}^{-1}$. The twisted exterior derivatives can be obtained from the ordinary exterior derivative using the adjoint action of the complex structures, as reviewed in our previous paper\mycite{deVriesSchroers:susyQMMM}: when $\mathfrak{d}$ is either $\d$, $\d_{\ICS_i}$ or a (twisted) Dolbeault operator, we have
\begin{align}
  \left[ \ad \ICS_i, \mathfrak{d} \right] & \ = \ \ICS_i \mathfrak{d} \ICS_i^{-1}\eqip{.} \anumber
\end{align}
Finally we also need that $\left[ \mathcal{L}_{\sogen_i} , \d \right] = 0$ and $\left[ \mathcal{L}_{\sogen_i} , \d_{\ICS_j} \right] = \sum_{k} \eps_{ijk} \d_{\ICS_k}$. We define raising and lowering operators as usual by $\Jang_{\pm} = \Jang_1 \pm i \Jang_2$, and we find that the algebra satisfied by the angular momentum operators and the Dolbeault operators is the following:
\begin{align}
  \left[ \Jang_3 , \partial \right] & \ = \ -\half \partial\eqip{,} &
  \left[ \Jang_3 , \partial_{\JCS} \right] & \ = \ \half \partial_{\JCS}\eqip{,} \nonumber \\
  \left[ \Jang_+ , \partial \right] & \ = \ -i \partial_{\JCS}\eqip{,} &
  \left[ \Jang_+ , \partial_{\JCS} \right] & \ = \ 0\eqip{,} \nonumber \\
  \left[ \Jang_- , \partial \right] & \ = \ 0\eqip{,} &
  \left[ \Jang_- , \partial_{\JCS} \right] & \ = \ i \partial\eqip{,} \anumber \\
\intertext{and}
  \left[ \Jang_3 , \overline{\partial} \right] & \ = \ \half \overline{\partial}\eqip{,} &
  \left[ \Jang_3 , \overline{\partial}_{\JCS} \right] & \ = \ -\half \overline{\partial}_{\JCS}\eqip{,} \nonumber \\
  \left[ \Jang_+ , \overline{\partial} \right] & \ = \ 0\eqip{,} &
  \left[ \Jang_+ , \overline{\partial}_{\JCS} \right] & \ = \ i \overline{\partial}\eqip{,} \nonumber \\
  \left[ \Jang_- , \overline{\partial} \right] & \ = \ -i \overline{\partial}_{\JCS}\eqip{,} &
  \left[ \Jang_- , \overline{\partial}_{\JCS} \right] & \ = \ 0\eqip{.} \anumber
\end{align}
Using the fact that $\mathfrak{d}^{\dag} = - * \overline{\mathfrak{d}} *$ and $[ \vec{\Jang} , * ] = 0$, we find
\begin{align}
  \left[ \Jang_3 , \partial^{\dag} \right] & \ = \ \half \partial^{\dag}\eqip{,} &
  \left[ \Jang_3 , \partial_{\JCS}^{\dag} \right] & \ = \ -\half \partial_{\JCS}^{\dag}\eqip{,} \nonumber \\
  \left[ \Jang_+ , \partial^{\dag} \right] & \ = \ 0\eqip{,} &
  \left[ \Jang_+ , \partial_{\JCS}^{\dag} \right] & \ = \ i \partial^{\dag}\eqip{,} \nonumber \\
  \left[ \Jang_- , \partial^{\dag} \right] & \ = \ -i \partial_{\JCS}^{\dag}\eqip{,} &
  \left[ \Jang_- , \partial_{\JCS}^{\dag} \right] & \ = \ 0\eqip{,} \anumber \\
\intertext{and}
  \left[ \Jang_3 , \overline{\partial}^{\dag} \right] & \ = \ -\half \overline{\partial}^{\dag}\eqip{,} &
  \left[ \Jang_3 , \overline{\partial}_{\JCS}^{\dag} \right] & \ = \ \half \overline{\partial}_{\JCS}^{\dag}\eqip{,} \nonumber \\
  \left[ \Jang_+ , \overline{\partial}^{\dag} \right] & \ = \ -i \overline{\partial}_{\JCS}^{\dag}\eqip{,} &
  \left[ \Jang_+ , \overline{\partial}_{\JCS}^{\dag} \right] & \ = \ 0\eqip{,} \nonumber \\
  \left[ \Jang_- , \overline{\partial}^{\dag} \right] & \ = \ 0\eqip{,} &
  \left[ \Jang_- , \overline{\partial}_{\JCS}^{\dag} \right] & \ = \ i \overline{\partial}^{\dag}\eqip{.} \anumber
\end{align}

We see that the Dolbeault operators $\overline{\partial}$ and $\partial_{\JCS}$ increase the total angular momentum of states (with respect to $\Jang_3$) by $\half$, while $\overline{\partial}_{\JCS}$ and $\partial$ decrease it by $\half$. The supercharges are therefore indeed spin-$\half$ operators.


\section{Example: charge-$1$ monopoles}
\label{sectEx1}
\subsection{Quantum mechanics on the  moduli space}
As a first example, we now consider a monopole of unit charge, in an $\nsusy=4$ supersymmetric Yang-Mills theory with maximal symmetry breaking. The following discussion applies to both the symmetry breaking $SU(2) \to U(1)$ and $SU(3) \to U(1) \times U(1)$. In the latter case, the charge-$1$ monopole is an embedding of the $SU(2)$ monopole in the $SU(3)$ model.
In this section and the next we use geometrical units, in which we have scaled the coupling constant and the vacuum expectation value of the Higgs field to unity, $e = 1$ and $a = 1$. This implies that Planck's constant $\hbar$ is dimensionless, but in general not equal to 1.

The moduli space of a single monopole in this theory is\mycite{GibbonsManton:CQDynBPSMon}
\begin{align}
\MS_1 \ = \ \RR^3 \times S^1\eqip{.} \anumber
\end{align}
The metric on $\MS_1$ is the flat metric, $\d s^2 = m \left( \d \vec{x}^2 \ + \ \d \chi^2 \right)$, where the mass of a monopole is determined by the Bogomol'nyi bound: $m = 4\pi$. The range of $\chi$ is $0 \leq \chi < 2\pi$. We define an orthonormal frame via
\begin{align}
  \ofr^i & \ = \ \sqrt{m} \, \d x^i\eqip{,} &
  \ofr^4 & \ = \ \sqrt{m} \, \d \chi\eqip{.} \label{vierbeinEx1}
\end{align}

For explicit calculations of the spin contents we will use the following
K\"ahler coordinates for the moduli space $\MS_1 \cong \CC \times \CC^*$\mycite{AtiyahHitchin}
\begin{align}
z^2 & \ = \ x^1 + ix^2\eqip{,} &
z^1 & \ = \ e^{x^3 + i \chi}\eqip{.} \label{kahlercoordM1}
\end{align}
We define the action of the complex structures $\ICS_i$ on 1-forms via
\begin{align}
\ICS_i (\ofr^j) & \ = \ \delta_{ij} \ofr^4 + \varepsilon_{ijk} \ofr^k\eqip{,} &
\ICS_i (\ofr^4) & \ = \ -\ofr^i\eqip{,} \label{complStrEx1}
\end{align}
where the vier-bein $e$ was defined in \eqref{vierbeinEx1}. The complex structure corresponding to the K\"ahler coordinates \eqref{kahlercoordM1} is $\ICS = \ICS_3$. A convenient basis of holomorphic 1-forms, with respect to this complex structure, is
\begin{align}
\alpha_2 & \ = \ \sqrt{\half[m]} \, \d z^2 \ = \ \frac{1}{\sqrt{2}} (\ofr^1 + i \ofr^2)\eqip{,} &
\alpha_1 & \ = \ \sqrt{\half[m]} \, \frac{\d z^1}{z^1} \ = \ \frac{1}{\sqrt{2}} (\ofr^3 + i \ofr^4)\eqip{.} \anumber
\end{align}
As we shall see below, the fermionic zero-modes can be conveniently described using this basis of holomorphic 1-forms. In terms of the K\"ahler coordinates and the basis of holomorphic 1-forms, the metric on $\MS_1$ can be written as
\begin{align}
\d s^2 & \ = \ m \left[ \left| \d z^2 \right|^2 + \left| \frac{\d z^1}{z^1} \right|^2 \right]
 \ = \ 2 |\alpha_2|^2 + 2 |\alpha_1|^2\eqip{.} \anumber
\end{align}
The action of the complex structures on the holomorphic 1-forms and their complex conjugates is also easily computed from equations \eqref{complStrEx1}. Using again the notation $\ICS= \ICS_3$ and $\JCS=\ICS_1$, we note the following, which will be useful later on.
\begin{align}
\ICS(\alpha_1) & \ = \ -i\alpha_1, &
\ICS(\alpha_2) & \ = \ -i\alpha_2, \nonumber \\
\JCS(\alpha_1) & \ = \ -i\overline{\alpha}_2, &
\JCS(\alpha_2) & \ = \ \phantom{-}i\overline{\alpha}_1. \label{IJonalpha}
\end{align}

The classical motion of the monopole in the moduli space approximation is given by the geodesics on $\MS_1$. Since $\MS_1$ is flat, these are straight lines, corresponding to uniform motion through space and a constant electric charge. The quantum mechanics of bosonic monopoles is described by wavefunctions ($0$-forms) on the moduli space, which must obey the Schr\"odinger equation,
\begin{align}
i \hbar \partial_t \Psi & \ = \ \half[\hbar^2] \Delta_{\MS_1} \Psi + m \Psi\eqip{,} &
\mbox{with} \quad  \Delta_{\MS_1} & \ = \
-\frac{\partial^2}{\partial\chi^2}- \frac{\partial^2}{\partial x_1^2}- \frac{\partial^2}{\partial x_2^2}
- \frac{\partial^2}{\partial x_3^2}\eqip{.} \label{SchrEq}
\end{align}

Starting with a solution $\Psi$ to the bosonic Schr\"odinger equation, we can use supersymmetry to find other solutions. We can generate the other states of the supermultiplet containing $\Psi$ by applying the supercharges, i.e. the Dolbeault operators $\partial$ and $\overline{\partial}$, their twisted counterparts $\partial_{\JCS}$ and $\overline{\partial}_{\JCS}$, and their adjoints $\partial^{\dag}$, $\overline{\partial}^{\dag}$, $\partial_{\JCS}^{\dag}$ and $\overline{\partial}_{\JCS}^{\dag}$. For a generic state $\Psi$, which is not an eigenstate of the  Laplace operator in the Schr\"odinger equation \eqref{SchrEq}, this leads to a supermultiplet consisting of $2^8=256$ states.

BPS states of supersymmetric monopoles are those states which have minimal energy for given  momentum and magnetic and electric charge. They correspond to short supermultiplets, made up of 16 states.  A short multiplet for a charge-$1$ monopole is obtained whenever the spin-0 state $\Psi$ with which we start is an eigenstate the Laplace operator. In order to lift the (infinite) degeneracy of (non-normalisable) bosonic eigenstates of the Laplace operator, we consider bosonic states of the form
\begin{align}
\Psi(\vec{x}, \chi) = e^{\frac{i }{\hbar} \vec{p}\cdot \vec{x}}e^{\frac{i }{\hbar} q \chi}, \anumber
\end{align}
which are also eigenstates of the momentum and electric charge operators, with eigenvalues $\vec{p}$ and $q$.  The states in a short multiplet containing such states can then be found using only the (twisted) Dolbeault operators $\partial$, $\overline{\partial}$, $\partial_{\JCS}$ and $\overline{\partial}_{\JCS}$.
For $\nsusy = 2$ supersymmetric monopoles, a short supermultiplet contains two singlets and one doublet under the angular momentum operator: the two singlets are $\Psi$ and $\overline{\partial} \overline{\partial}_J \Psi$; the doublet is $\left\{\overline{\partial} \Psi, \overline{\partial}_J \Psi\right\}$.
The short multiplet of a monopole in the $\nsusy = 4$ supersymmetric theory decomposes as follows:
\begin{align}
\mbox{5 singlets:} & \qquad \Psi, \ \overline{\partial}_J \overline{\partial} \Psi, \ \partial_J \partial \Psi, \ \partial_J \partial \overline{\partial}_J \overline{\partial} \Psi, \ \frac{1}{\sqrt{2}} \left( \partial_J \overline{\partial} - \partial \overline{\partial}_J \right) \Psi \nonumber \\
\mbox{4 doublets:} & \qquad \left\{\overline{\partial} \Psi, \,  \overline{\partial}_J \Psi\right\}, \ \left\{ \partial \Psi, \,  \partial_J \Psi \right\}, \ \left\{ \overline{\partial} \partial_J \partial \Psi, \,  \overline{\partial}_J \partial_J \partial \Psi \right\}, \ \left\{ \partial \overline{\partial}_J \overline{\partial} \Psi, \,  \partial_J \overline{\partial}_J \overline{\partial} \Psi \right\} \nonumber \\
\mbox{1 triplet:} & \qquad \left\{ \partial \overline{\partial} \Psi, \,  \frac{1}{\sqrt{2}} \left( \partial_J \overline{\partial} + \partial \overline{\partial}_J \right) \Psi , \,  \partial_J \overline{\partial}_J \Psi \right\} \nonumber
\end{align}
It is not difficult to compute these states explicitly. The action of $\partial_J$, for example, can be computed using its definition in terms of $\overline{\partial}$ and $\JCS$ given in equations \eqref{defTwistDolb} and by applying $\JCS$ according to equations \eqref{IJonalpha}.

If one computes states in the multiplet containing the bosonic state $\Psi=\exp(\frac{i}{\hbar} q\chi)$ in this way, one finds expressions which are simple linear combinations (with constant coefficients) of wedge products of  the forms $\alpha_1$, $\alpha_2$ and their complex conjugates, multiplied with the original function $\Psi=\exp(\frac{i}{\hbar} q\chi)$. We compute these linear combinations in the following section using a different method.
In terms of the original field theory, the states one obtains  describe bound states of a dyon and  Majorana fermions. More precisely,  they correspond to  a magnetic monopole at rest carrying electric charge $q$, while each $ \overline{\alpha}_n$, $n=1$ or $2$, indicates the occupation of a Dirac zero-mode $\xi^+$ and each  $\alpha_n$, $n=1$ or $2$, indicates the occupation of a Dirac zero-mode $\xi^-$ (where we used the quantisation prescription for fermionic zero-modes given in equations \eqref{fermquant}).

\subsection{Angular momentum and spin}
\label{sectEx1Spin}
For the case of a single monopole,  the vector fields $\sogen_i$  appearing in  total angular momentum operator $\vec{\Jang}$  \eqref{Jang} can be written explicitly as
\begin{align}
\sogen_i & \ = \ -\varepsilon_{ijk} x^j \partial_k\eqip{,} &
\left[\sogen_i, \sogen_j\right] & \ = \ \varepsilon_{ijk} \sogen_k\eqip{.} \anumber
\end{align}
One checks that the Lie derivatives with respect to these vector fields obey  the commutation relations $\left[\mathcal{L}_{\sogen_i}, \ad \ICS_j\right] = \varepsilon_{ijk} \ad \ICS_k$ with the complex structures.


%
We can write the Lie-derivative in terms of the exterior derivative and interior product using Cartan's formula,
\begin{align}
\mathcal{L}_{\sogen_i} & \ = \ \iota_{\sogen_i} \d + \d \iota_{\sogen_i}\eqip{.} \anumber
\end{align}
The complex structure $\ad \ICS_i$ and the term $\d \iota_{\sogen_i}$ act trivially on functions, as $0$. Furthermore, since the 1-forms $\alpha_1$, $\alpha_2$, $\overline{\alpha}_1$ and $\overline{\alpha}_2$ are closed, the term $\iota_{\sogen_i} \d$ acts on them as $0$. This suggests that we identify the orbital angular momentum and spin operators as
\begin{align}
L_i & \ = \ i \left( \iota_{\sogen_i} \d \right)\eqip{,} &
S_i & \ = \ i \left( \d \iota_{\sogen_i} - \half \ad \ICS_i \right)\eqip{,} \label{defLSop}
\end{align}
which both act on forms obeying the Leibniz rule.
With these definitions,
\begin{align}
  \vec{L}(\alpha_n) \ = \ \vec{L}(\overline{\alpha}_n) \ = \ & 0,  \quad n \in \{1,2\}\eqip{,}
\end{align}
and  $\vec{S}(f) = 0 $ for any function $f$. We therefore find that the spin operator acts on $\alpha_1$, $\alpha_2$, $\overline{\alpha}_1$ and $\overline{\alpha}_2$  according to
\begin{subequations}\begin{align}
\Jang_i(\alpha_m) \ = \ S_i(\alpha_m) & \ = \ -\half (\overline{\pauli}_i)_{mn} \alpha_n\eqip{,} \anumber \\
\Jang_i(\overline{\alpha}_m) \ = \ S_i(\overline{\alpha}_m) & \ = \ \phantom{-}\half (\pauli_i)_{mn} \overline{\alpha}_n\eqip{.} \anumber
\end{align}\end{subequations}
The pair $\left\{\alpha_1, \alpha_2\right\}$ form the two-dimensional conjugate representation to the pair $\left\{\overline{\alpha}_1, \overline{\alpha}_2\right\}$. These two representations are isomorphic, as may be seen by applying the
the unitary transformation
\begin{align}
  \beta_1 & \ = \ -i\alpha_2\eqip{,} &
  \beta_2 & \ = \ i\alpha_1\eqip{,} \anumber
\end{align}
and checking that
\begin{align}
\Jang_i(\beta_m) \ = \ S_i(\beta_m) & \ = \ \half (\pauli_i)_{mn} \beta_n\eqip{.} \anumber
\end{align}

Decomposing  the full $\nsusy = 4$ supermultiplet of forms  into irreducible representations of the spin operator, we find  the following five  singlet states
\begin{align}
  1, &&
  \overline{\alpha}_2 \wedge \overline{\alpha}_1, &&
  \alpha_1 \wedge \alpha_2, &&
  \frac{i}{\sqrt{2}}  \left( \alpha_1 \wedge \overline{\alpha}_1 + \alpha_2 \wedge \overline{\alpha}_2 \right), &&
  \alpha_1 \wedge \alpha_2 \wedge \overline{\alpha}_2 \wedge \overline{\alpha}_1.
\end{align}
We have two doublets given by
\begin{align}
 \{\overline{\alpha}_1, \, \overline{\alpha}_2 \}, && \{-i\alpha_2, \, i\alpha_1\},
\end{align}
and we obtain another two doublets, dual to the first two, by applying the Hodge-star operator:
\begin{align}
  \{\alpha_1 \wedge \alpha_2 \wedge \overline{\alpha}_1 , \, \alpha_1 \wedge \alpha_2 \wedge \overline{\alpha}_2\}, &&
  \{-i \alpha_2 \wedge \overline{\alpha}_2 \wedge \overline{\alpha}_1, \,
   i \alpha_1 \wedge \overline{\alpha}_2 \wedge \overline{\alpha}_1 \}.
\end{align}
Finally, there is  triplet, consisting of the following states
\begin{align}
\{-i \alpha_2 \wedge \overline{\alpha}_1, \, \frac{1}{\sqrt{2}} i \left( \alpha_1 \wedge \overline{\alpha}_1 - \alpha_2 \wedge \overline{\alpha}_2 \right) , \,  i \alpha_1 \wedge \overline{\alpha}_2\}.
\end{align}
We see that the 16-dimensional space of 1-forms decomposes into  five singlets, four  doublets  and one triplet for our spin operator --  in perfect agreement with Osborn's result for the zero-modes of the $\nsusy = 4$ supersymmetric monopoles, as reviewed after equation \eqref{spinOpFT}.

It is interesting to relate the states in the various multiplets to other geometrically interesting forms.
The 2-forms which are singlets are all anti-self-dual, and related to the hyperk\"ahler forms $\omega_i$, defined by $\omega_i(X,Y) = g(X, \ICS_i Y)$, as follows
\begin{align}
\omega_1 &\  = \ \ofr^4 \wedge \ofr^1 - \ofr^2 \wedge \ofr^3 \ = \ -i  (\overline{\alpha}_2 \wedge \overline{\alpha}_1+ \alpha_1 \wedge \alpha_2 )\eqip{,} \nonumber \\
\omega_2 &  \ = \  \ofr^4 \wedge \ofr^2 - \ofr^3 \wedge \ofr^1 \ = \ \phantom{-i (}\overline{\alpha}_2 \wedge \overline{\alpha}_1- \alpha_1 \wedge \alpha_2 \phantom{)}\eqip{,} \nonumber \\
\omega_3 & \ = \ \ofr^4 \wedge \ofr^3 - \ofr^1 \wedge \ofr^2  \ = \ -i ( \alpha_1 \wedge \overline{\alpha}_1 + \alpha_2 \wedge \overline{\alpha}_2 )
\eqip{.}
\end{align}
By contrast, the 2-forms corresponding to the triplet states are self-dual.  They can be combined into the three forms
\begin{align}
\label{selfdualforms}
T^1 & \ =  \ \ofr^4 \wedge \ofr^1 + \ofr^2 \wedge \ofr^3 \ = \ -i(\alpha_1\wedge \overline{\alpha}_2 + \alpha_2 \wedge \overline{\alpha}_1 )  \eqip{,} \nonumber \\
T^2 & \ = \ \ofr^4 \wedge \ofr^2 + \ofr^3 \wedge \ofr^1  \ = \ \phantom{-i (}\alpha_1\wedge \overline{\alpha}_2 - \alpha_2 \wedge \overline{\alpha}_1\phantom{)}\eqip{,} \nonumber \\
T^3 & \ =  \ \ofr^4 \wedge \ofr^3 + \ofr^1 \wedge \ofr^2 \ = \ -i(\alpha_1\wedge\overline{\alpha}_1 -\alpha_2\wedge \overline{\alpha}_2)\eqip{,}
\end{align}
which are all of type $(1,1)$ with respect to the complex structure $\ICS$ and satisfy
\begin{align}
S_i T^j & \ = \ i \varepsilon_{ijk} T^k\eqip{.} \anumber
\end{align}
Finally, the canonical holomorphic symplectic 2-form $\Omega_3$ corresponding to the complex structure $\ICS$ is the following singlet:
\begin{align}
  \Omega_3  \ = \ \omega_1 + i \omega_2 \ = \ -2i  \alpha_1 \wedge \alpha_2.
 \end{align}


\section{Example: charge-$(1,1)$ monopoles}
\label{sectEx2}
\subsection{The geometry of the moduli space and a set of K\"ahler coordinates}
As a second example we study charge-$(1,1)$ monopoles in $\nsusy=4$ supersymmetric Yang-Mills theory with
symmetry breaking $SU(3) \to U(1) \times U(1)$. These monopoles may be thought of as being composed of two constituent monopoles which are each $SU(2)$ BPS monopoles of charge 1, but embedded into $SU(2)$ subgroups associated with different simple roots of $SU(3)$. The masses of the constituents depend on the direction of the vacuum expectation value of the Higgs field in the Cartan subalgebra of $SU(3)$ and are denoted $m_1$ and $m_2$ in the following.

The main references for the magnetic monopoles of charge-$(1,1)$ studied in this section are the papers by Gauntlett and Lowe\mycite{GauLowe:DyonsSDuality}, and by Lee, Weinberg and Yi\mycite{LeeWeinbergYi:EMDSU3Mon};
they include a detailed discussion of magnetic and electric charges and a derivation of the metric on moduli space for charge-$(1,1)$ monopoles, which is
\begin{align}
  \MS_{1,1} & \ = \ \RR^3 \times \frac{\RR \times \MTN}{\ZZ}\eqip{,} \label{11mopomod}
\end{align}
where $\MTN$ is the 4-dimensional Taub-NUT manifold with a positive length parameter. As we discussed in our previous paper\mycite{deVriesSchroers:susyQMMM}, the centre of mass dynamics and the relative motion can be separated. $\MTN$ is the factor of the moduli space corresponding to the relative motion. Topologically $\MTN \cong \RR^4$, but it has a curved metric, given below.

For practical calculations it is usually convenient to work with the covering space of the moduli space, $\widetilde{\MS}_{1,1} = \RR^3 \times \RR \times \MTN$, and impose the identification by $\ZZ$ on the results at the end. The physics behind the $\ZZ$ action is explained carefully in the main references\mycite{GauLowe:DyonsSDuality,LeeWeinbergYi:EMDSU3Mon}, and also summarised in our earlier paper\mycite{deVriesSchroers:susyQMMM}

The metric on the centre of mass moduli space $\RR^3 \times \RR$ is the flat metric. Using centre of mass coordinates $\vec{R}$ and $\chi$, the metric is analogous to that of the single monopole moduli space $\MS_1$. The line element is  $\d s^2 = M \left( \d \vec{R}^2 \ + \ \d \chi^2 \right)$, where $M$ is the total mass of the charge-$(1,1)$ monopole.

The manifold $\MTN \cong \RR^4$ can be coordinatised in terms of a radial coordinate $r$ and
Euler angles $\theta$, $\phi$ and $\psi$ on $S^3$, with the ranges $0 \leq \theta < \pi$, $0 \leq \phi < 2\pi$ and $0 \leq \psi < 4\pi$. As explained in the main references\mycite{GauLowe:DyonsSDuality,LeeWeinbergYi:EMDSU3Mon}, the angle $\psi$ is the conjugate variable to half the difference between the electric charges of the constituent monopoles. The range $[0,4\pi)$ reflects the fact that half this difference necessarily is an element of $\half \ZZ$.
Finally, the division by $\ZZ$ on the total moduli space corresponds to identifying the points
\begin{align}
(\vec{R}, \chi, \vec{r}, \psi) & \ \sim \ (\vec{R}, \chi+2\pi, \vec{r}, \psi+\tfrac{4m_2}{m_1+m_2}\pi)\eqip{,} \label{Z2identificationM11}
\end{align}
which, as mentioned above, depends on the masses of the constituent monopoles\mycite{GauLowe:DyonsSDuality,LeeWeinbergYi:EMDSU3Mon}.

The metric on the Taub-NUT manifold $\MTN$ can be expressed in terms of the vector $\vec{r} = (x_1, x_2, x_3)$, which gives the relative position of the two monopoles and is related to the radial coordinate $r$ and the Euler angles $\theta$ and $\phi$ via
\begin{align}
x_1 & \ = \ r \sin\theta \cos\phi\eqip{,} &
x_2 & \ = \ r \sin\theta \sin\phi\eqip{,} &
x_3 & \ = \ r \cos\theta\eqip{,} \nonumber
\end{align}
and the right-invariant 1-forms
\begin{align}
 \eta_1 & \ = \ - \sin \psi \d \theta + \cos \psi \sin \theta \d \phi\eqip{,} \nonumber \\
 \eta_2 & \ = \ \phantom{-} \cos \psi \d \theta + \sin \psi \sin \theta \d \phi\eqip{,} \nonumber \\
 \eta_3 & \ = \ \phantom{-} \, \d \psi + \cos \theta \d \phi \eqip{,} \anumber
\end{align}
which satisfy $\d \eta_i = \half \eps_{ijk} \eta_j \wedge \eta_k$.
The metric is given by
\begin{align}
\d s^2 & \ = \ \mu \Big[ V ( \d \vec{r} \cdot \d \vec{r} ) + V^{-1} (\eta_3)^2 \Big] \nonumber \\
 & \ = \ \mu \Big[ V \left(\d r^2 + r^2 \left( (\eta_1)^2 + (\eta_2)^2 \right) \right) + V^{-1} (\eta_3)^2 \Big]\eqip{,} \label{taubnutmetric}
\end{align}
where $\mu$ is the reduced mass of the monopole system, and
\begin{align}
 V \ = \ & \left(1 + \frac{1}{r}\right)\eqip{.}
\end{align}
It will be convenient to describe the metric in terms of the following vier-bein:
\begin{align}
 e^i & \ = \ \sqrt{\mu V} \d x^i, &
 e^4 & \ = \ \sqrt{ \frac{\mu}{V} }\eta_3.
\end{align}
A set of K\"ahler coordinates on the Taub-NUT manifold $\MTN$ is defined by\mycite{GibRub:HiddenSymMM}
\begin{align}
w & \ = \ r \sin \theta e^{i\phi}
\eqip{,} &
v & \ = \ r (1+\cos\theta) e^{r \cos \theta +i(\psi+\phi)}\eqip{.} \anumber
\end{align}
We define the 1-forms $\alpha_2$ and $\alpha_1$, which form a convenient basis of holomorphic 1-forms with respect to the complex structure corresponding to the K\"ahler coordinates $w$ and $v$,  via
\begin{subequations}\begin{align}
\alpha_2 & \ = \ \frac{1}{\sqrt{2}} (e^1 + i e^2)
  \ = \ \sqrt{\frac{\mu V}{2}} \ \d w\eqip{,} \anumber \\
\alpha_1 & \ = \ \frac{1}{\sqrt{2}} (e^3 + i e^4)
  \ = \ \sqrt{\frac{\mu}{2V}} \left( \frac{\d v}{v} + (\cos\theta - 1) \frac{\d w}{w} \right)\eqip{.} \anumber
\end{align}\end{subequations}
The Taub-NUT metric can then be written in terms of the K\"ahler coordinates as
\begin{align}
\d s^2 & \ = \ \mu \left[ V \left| \d w \right|^2 + V^{-1} \left| \frac{\d v}{v} + (\cos \theta - 1) \frac{\d w}{w} \right|^2 \right]
 \ = \ 2 |\alpha_2|^2 + 2 |\alpha_1|^2\eqip{.} \anumber
\end{align}

The classical dynamics are described by the geodesics on the moduli space. The centre of mass dynamics, corresponding to motion in $\RR^3 \times \RR$, is analogous to the single monopole dynamics discussed in example 1. For the relative motion of the two monopoles, described by the moduli space $\MTN$, we found in our previous paper\mycite{deVriesSchroers:susyQMMM} that there are no bound orbits, and hence there are only scattering solutions for the charge-$(1,1)$ monopole.
Similarly, the quantum mechanics of  bosonic charge-$(1,1)$ monopoles, governed by the Schr\"odinger equation on the moduli space, only gives scattering solutions; bound states of  bosonic monopoles do not exist in this case.

\subsection{$\nsusy = 4$ supersymmetric  quantum mechanics of charge-$(1,1)$ monopoles}
\label{sectEx2N2}
The product structure of the moduli space allows us to separate centre of mass and relative motion
in the quantum theory. A general form on the moduli space can be written as a linear combination of wedge products of forms on the centre of mass and relative moduli spaces. The supercharges, the (twisted) Dolbeault operators and their adjoints, decompose into a sum of (twisted) Dolbeault operators on the centre of mass and relative moduli spaces. The Laplacian, too, decomposes into the sum of centre of mass and relative moduli space components. For a form $\upsilon =\upsilon_1 \wedge \upsilon_2$, where $\upsilon_1$ and $\upsilon_2$ are forms on $\RR^3 \times \RR$ and $\MTN$ respectively,
\begin{align}
  \Delta_{M_{1,1}}\upsilon = ( \Delta_{\RR^3 \times \RR} \upsilon_1 ) \wedge \upsilon_2 + \upsilon_1 \wedge ( \Delta_{\MTN} \upsilon_2 )\eqip{.}
\end{align}

Focussing on the moduli space for the relative motion of the monopoles, we can generate multiplets of states, starting with a wavefunction $\Phi$ on $\MTN$, by applying the (twisted) Dolbeault operators on the Taub-NUT manifold. When $\Phi$ is an eigenstate of the Laplacian $\Delta_{\MTN}$ we obtain, in general, 16 independent states in the $\nsusy = 4$ supersymmetric model. By taking the wedge product of these states with a multiplet of centre of mass states, we obtain multiplets of 256 states on the total moduli space, all with the same energy.


For the remainder of this section we concentrate on BPS states in the original field theory, which  correspond to short multiplets of 16 states on the total moduli space. Since the short multiplet on the centre of mass moduli space already involves 16 states, BPS states require the existence of a  unique normalisable  form on the relative moduli space which is annihilated by all the supercharges. Such a form is necessarily  either self-dual or anti-self dual (otherwise its Hodge dual would provide a second state of the same energy)  and harmonic.
We call such a form  a Sen-form in the remainder of this paper, since its existence and relevance for checking S-duality was first established, in the context of $SU(2)$ monopoles, by A.~Sen\mycite{Sen:DMBS}.  As has been shown before\mycite{GauLowe:DyonsSDuality,LeeWeinbergYi:EMDSU3Mon}, such a form does exist on the Taub-NUT manifold. In the following, we shall exhibit some of its properties explicitly and, in particular, analyse its behaviour under our angular momentum operator.

The Sen form on the Taub-NUT manifold is given by
\begin{align}
\omega_S & \ = \ \frac{r}{r+1} \eta_1 \wedge \eta_2 + \frac{1}{(r+1)^2} \d r \wedge \eta_3 \ = \ \d (V \eta_3) \label{TNSenform}
\end{align}
It is normalisable, since $\int \omega_S \wedge \omega_S = 8\pi^3 \int \frac{r}{(r+1)^3} \d r = 4\pi^3$ is finite. It can be rewritten as
\begin{align}
\omega_S 
 & \ = \ - \frac{1}{\mu} V^{-2} \left( \partial_i V \right) T^i \label{TNSenforma}
\end{align}
where we have defined the triplet $T^i$ as in \eqref{selfdualforms} by $T^i = e^4 \wedge e^i + \half \eps_{ijk} e^j \wedge e^k$. Since the 2-forms $T^i$ are self-dual, the Sen-form is self-dual as well.

One can read off directly from the expression for self-dual forms \eqref{selfdualforms} on the single monopole moduli space that these forms are of degree $(1,1)$ with respect to the complex structure $\ICS$.  A similar analysis on $\MTN$ shows that the Sen form  is of degree $(1,1)$
(in fact, it is of degree $(1,1)$ with respect to any of the complex structures) and therefore not an anti-holomorphic state corresponding to any fermionic or bosonic state of the $\nsusy = 2$ supersymmetric system. Unlike the $\nsusy=4$ supersymmetric model, the $\nsusy=2$ supersymmetric charge-$(1,1)$ monopole system therefore has no BPS states in the moduli space approximation.

The total angular momentum operator $\vec{\Jang}$ is once again defined by equation \eqref{Jang}. As usual we decompose the total moduli space into the centre of mass and relative moduli spaces. The vector fields generating the $SO(3)$ action on the Taub-NUT manifold  are denoted  $\xi^L_i$ (see also Gibbons and Manton\mycite{GibbonsManton:CQDynBPSMon}). They are given by
\begin{subequations}\begin{align}
\xi^L_1 \ = \ & - \frac{\cos\phi}{\sin\theta} \frac{\partial}{\partial  \psi} + \sin\phi
\frac{\partial} {\partial \theta} + \frac{\cos\theta}{\sin\theta} \cos\phi \frac{\partial}{\partial \phi} \anumber \\
\xi^L_2 \ = \ & - \frac{\sin\phi}{\sin\theta} \frac{\partial}{\partial \psi} - \cos\phi
\frac{\partial} { \partial \theta} + \frac{\cos\theta}{\sin\theta} \sin\phi\frac{\partial} {\partial \phi} \anumber \\
\xi^L_3 \ = \ & - \frac{\partial}{\partial \phi} \anumber
\end{align}\end{subequations}
and satisfy
\begin{align}
  \left[\xi^L_i, \xi^L_j\right] & \ = \ \varepsilon_{ijk} \xi^L_k\eqip{.} \anumber
\end{align}
Again, the Lie derivatives with respect to these vector fields obey the following commutation relations with the complex structures: $\left[\mathcal{L}_{\xi^L_i}, \ad \ICS_j\right] = \varepsilon_{ijk} \ad \ICS_k$. We find that the total angular momentum operator acts on $\overline{\alpha}_1$ and $\overline{\alpha}_2$ as
\begin{subequations}\begin{align}
\Jang_i(\alpha_m) & \ = \ -\half (\overline{\pauli}_i)_{mn} \alpha_n\eqip{,} \anumber \\
\Jang_i(\overline{\alpha}_m) & \ = \ \phantom{-}\half (\pauli_i)_{mn} \overline{\alpha}_n\eqip{.} \anumber
\end{align}\end{subequations}
Once more, we have that $\{\overline{\alpha}_1, \overline{\alpha}_2\}$ and $\{ -i \alpha_2, i \alpha_1 \}$ form doublets under the angular momentum operator, and the multiplet decomposition on the Taub-NUT manifold is the same as that on the flat moduli space in section \ref{sectEx1Spin}.

As explained earlier,  the Sen form needs to be the unique harmonic normalisable form on the relative moduli space in order to  be part of a short supersymmetry multiplet. In particular, we therefore expect it to be a singlet   of the total angular momentum operator. It well-kown (and easy to check) that the Sen form is invariant under the geometric $SO(3)$ action on the moduli space. In fact it is a straightforward calculation to show that the Sen form is a singlet under our angular momentum operator as well: it is of the form $\omega_S = U_j T^j$ for a vector $U_j$ so that
\begin{align}
  \Jang_i \omega_S & \ = \ \Jang_i (U_j T^j) \ = \ (\eps_{ijk} U_k) T^j + U_j (\eps_{ijk} T^k) \ = \ 0\eqip{,} \anumber
\end{align}
as required.


\section{Outlook}
\label{outlook}
The main motivation for studying quantised monopole dynamics over the last 15 years or so has been to test the S-duality conjecture, which generalises the  Montonen-Olive conjecture\mycite{MontOlive:MMGP} by incorporating both supersymmetry and dyonic states.  Almost all the tests have been related to establishing the existence of BPS states predicted by S-duality and, so far, they have all supported the conjecture. If S-duality holds,  the strongly coupled physics of massive, electrically charged particles (like W-bosons) can be studied in terms of the physics of magnetic monopoles at weak (electric) coupling.  However, in order to carry out such a study in practice, one needs to be able to say which quantum states on the electric side are dual to which quantum states on the magnetic side.  The only practical way of doing this is to characterise states in terms of quantum numbers like electric and magnetic charge, angular momentum and spin,  and  to exploit their supersymmetry multiplet structure. Our formulae for the supercharges and the angular momentum operator as differential operators make it possible to  organise quantum states of magnetic monopoles in this way.
They are, therefore,  likely to be  essential both   in more detailed tests and in applications of the S-duality conjecture like the ones sketched in the outlook section of our previous paper\mycite{deVriesSchroers:susyQMMM}. Obvious candidates for carrying out such calculations are the theory with gauge group $SU(3)$ broken to $U(1) \times\ U(1)$,  on which we focussed in this paper, but also the theory with   $SU(2)$ broken to $U(1)$. In the latter  case, the moduli space for the relative motion of two monopoles is the Atiyah-Hitchin manifold. The bosonic quantum mechanics on the Atiyah-Hitchin manifold was studied by one of us\mycite{Schroers:QSBPSMon}, building on earlier work by Gibbons and Manton\mycite{GibbonsManton:CQDynBPSMon}. The  quantum mechanics of $N=4$ supersymmetric  monopoles can thus be computed, at least in principle,  by  using the expression for the complex coordinates found by Olivier\mycite{Olivier:coomplex} and applying the expressions for the supercharges given in this paper.


\section*{Acknowledgements}
EJdV is grateful to Jos\'e Figueroa-O'Farrill for useful discussions on the quantisation and supersymmetries of the effective action.




\markboth{\refname}{\refname}
\bibliography{bib}


\end{document}